\documentclass[12pt]{iopart}
\usepackage{graphicx}
\usepackage{dcolumn}
\usepackage{bm}
\usepackage{iopams}
\usepackage{setstack}

\newcommand{\ket}[1]{\left| #1 \right \rangle }

\newcommand{\hc}{\hat{c}^{\phantom{\dagger}}}
\newcommand{\hcd}{\hat{c}^{\dagger}}

\newcommand{\langind}[1]{\hspace{-0.1cm}{\phantom{\rangle}}_{#1}
\hspace{-0.05cm} \langle\hspace{0.03cm} }  
\newcommand{\rangind}[1]{\rangle\hspace{-0.2cm}{\phantom{\rangle}}_{#1}
\hspace{0.05cm}}
\newcommand{\qvec}{{\bf q}} 
 
\newcommand{\kvec}{{\bf k}} 

\begin{document}

\title{Linear-Response Dynamics from the Time-Dependent Gutzwiller Approximation} 

\author{J.~B\"unemann$^{1,2}$, M.~Capone$^3$, J.~Lorenzana$^{4}$, 
and G.~Seibold$^2$}

\address{$^1$Max-Planck-Institute for Solid-State Research, Heisenbergstr.~1, 70569 Stuttgart, Germany}

\address{$^2$Institut f\"ur Physik, BTU Cottbus, P.O. Box 101344, 03013 Cottbus, Germany}

\address{$^3$ CNR-IOM-Democritos National Simulation Centre and International School
for Advanced Studies (SISSA), Via Bonomea 265, I-34136, Trieste, Italy}

\address{$^4$ ISC-CNR and
Dipartimento di Fisica, Universit\`a di Roma ``La Sapienza'', Piazzale Aldo
Moro 5, I-00185 Roma, Italy}

\ead{$^1$buenemann@gmail.com}

\begin{abstract}%
Within a Lagrangian formalism we derive the time-dependent 
Gutzwiller approximation for general multi-band Hubbard models.
Our approach explicitly incorporates the coupling between time-dependent
variational parameters and a time-dependent density matrix from which 
we obtain dynamical correlation functions in the linear response regime.   
Our results are illustrated for the one-band model
where we show that the interacting system can be mapped to an effective
problem of fermionic quasiparticles coupled to ``doublon'' (double occupancy) bosonic
fluctuations. The latter have an energy on the scale of the on-site
Hubbard repulsion $U$ in the dilute limit but becomes soft
at the Brinkman-Rice transition which is shown to be related to an
emerging conservation law of doublon charge and the associated gauge
invariance. Coupling with the boson mode produces structure in the
charge response and we find that a similar structure appears in  
dynamical mean-field theory. 
\end{abstract}

 \submitto{\NJP}
        
\today
  
\maketitle 
             
\section{Introduction}
Recent advances in ultra-fast spectroscopy allow us to monitor
the dynamics of electrons on a femtosecond scale.      
This is especially interesting for strongly correlated materials, 
such as high-temperature superconductors \cite{conte,fausti,mansart},      
since in their case the spectroscopic probe is able to investigate
the intra-electronic redistribution of excitation energies before
the relaxation via the lattice starts. From the theoretical point of view, 
 this is obviously a challenging problem since it requires a method capable of treating
the relaxation dynamics of a strongly correlated system out of equilibrium.
In this regard, a state-of-the art approach is the dynamical mean-field theory 
(DMFT) which 
has recently been applied \cite{eckstein} 
to the single-band Hubbard model in order to study the
double-occupancy relaxation after laser excitation. 
However, for the application to real systems this method 
is rather demanding from a numerical                                       
point of view since it requires the self-consistent solution of         
complex single-impurity models driven out of equilibrium.

In this regard, the time-dependent Gutzwiller approximation (TDGA) is a
promising alternative since it joins the numerical simplicity 
of standard random phase approximation (RPA) 
with the ability to
capture important many-particle effects, as the Mott-Hubbard
transition. In a series of papers 
\cite{goe01,goe03,goe042,goe08},  
we have developed the TDGA which is
based on a variational Ansatz for the Hubbard model \cite{GUTZ1,GUTZ2}
evaluated in the limit of infinite spatial dimensions \cite{GEBHARD}. 
This approach has recently been generalised for the study of
multi-band Hubbard models \cite{oelsen1,oelsen2} and is based
on the expansion of the Gutzwiller energy functionals which depend
on the density matrix and variational parameters related to the
atomic eigenstates.                 
In previous work \cite{goe01,goe03}
the latter have been eliminated
by assuming that they instantaneously adjust to the density fluctuations
(`antiadiabaticity approximation'). As a result, one obtains an energy       
functional which only depends on the density matrix and therefore       
the RPA  for density-dependent forces \cite{ring,blaizot} 
can be applied in order to evaluate response functions.                        
This (approximate) version of the TDGA has been applied successfully
 to the evaluation of 
dynamical correlation      
functions in cuprate superconductors \cite{goe12} including the optical conductivity         
\cite{lor03} and the magnetic susceptibility \cite{goetz05,sei06}.             
It has also been related to Auger spectroscopy by calculating pair               
excitations in one- \cite{goe08} and three-band \cite{ugenti10} Hubbard models.             

More recently the TDGA was extended by Schir\'{o} and 
Fabrizio~\cite{schiro1,schiro2,fabrizio}                          
towards the inclusion of time-dependent variational parameters.        
Concerning the evaluation of response functions this approach           
can supersede the `antiadiabaticity assumption', mentioned  
above, since the double-occupancy dynamics follows from a time-dependent
variational principle. However, in  Refs.~\cite{schiro1,schiro2} 
the authors focused on the study of quantum quenches for systems
with a homogeneous ground state. In this case, the time dependence
is captured by the double-occupancy dynamics whereas the
single-particle density matrix is time-independent. 
Recent developments consider simultaneously the dynamics of the 
 double occupancy and of the density matrix~\cite{andre,mazza}.

In this paper we will re-derive the TDGA for a time-dependent
Gutzwiller variational wave-function applied to multi-band
Hubbard models. Our resulting equations of motion will explicitly
capture the coupling between the time-dependent variational
parameters and the density matrix. We will analyse these equations
in the small-amplitude (i.e., linear response) limit 
and apply the theory to the evaluation of dynamical charge
correlations in the single-band case.
It turns out that the previous formulation of the TDGA~\cite{goe01,goe03}
is recovered in the low-frequency limit. However, the incorporation
of fluctuations in the time-dependent density of double occupied
states (``doublons'')  leads to additional
spectral weight above the band-like excitations in very good
agreement with exact diagonalisation and DMFT.

The paper is organised as follows: In Sec.~\ref{sec:vp} we introduce
the time-dependent variational principle which is underlying the present
work. Since the corresponding expectation values are evaluated with 
multi-band Gutzwiller wave-functions the latter are presented in 
Sec.~\ref{sec2}. The evaluation of time-dependent matrix elements
is performed in Sec.~\ref{secmat} which allows for the
derivation of the Lagrangian and corresponding equations of
motion in Sec.~\ref{df67c}.
Our investigations are specified for the single-band Hubbard model
in Sec.~\ref{sec:oneb}, where we also discuss a two-site example
which can be treated analytically.
Finally, the small amplitude limit of the TDGA is derived in
Sec.  \ref{sec:sma} and discussed in the context of response
functions in Sec.~\ref{hjud}.
Numerical results for the dynamical
charge susceptibility are presented in Sec.~\ref{sec:res} and compared 
with dynamical
mean-field theory (DMFT) and exact diagonalisation.
We finally conclude our investigations in Sec.~\ref{sec:con}.

\section{The time-dependent Gutzwiller theory}\label{sec1}

\subsection{Variational Principle}\label{sec:vp}


The time-dependent Schr\"odinger equation for a general time-dependent
Hamiltonian $\hat{H}(t)$
 ($\hbar=1$),
can be obtained by requesting that the action 
\begin{equation}
  \label{eq:action}
  S=\int dt L(t)
\end{equation}
is stationary with respect to variations of the wave function. 
It is in general convenient to perform this variation based
on a real Lagrangian \cite{kramers} 
\begin{equation}
  \label{eq:lagran}
  L(t)=\frac{\rmi}{2}\frac{\langle \Psi|\dot{\Psi}\rangle
-\langle \dot{\Psi}|\Psi\rangle} 
{ \langle \Psi | \Psi \rangle}
- \rmi\frac{\langle \Psi|\hat{H}|\Psi\rangle} 
{ \langle \Psi | \Psi \rangle}\;,
\end{equation}
which leads to equations of motion that are independent of the
phase and the norm of the wave-functions.

If one restricts the wave-function $\ket{\Psi( z_i(t)) }$ to a 
certain trial form, depending on a set of (in general complex) `variational parameters' 
$z_i(t)$, the Euler-Lagrange equations for $z_i(t)$ provide 
 an approximation for the time evolution of the system. 
  For example, restricting the wave-function 
to Slater determinants and using the amplitude of the
single particle orbitals as variational parameters yields the
time-dependent Hartree-Fock approximation.  In this work, we will consider 
variational wave
functions of the Gutzwiller form \cite{GUTZ1,GUTZ2,GEBHARD}
for general multi-band Hubbard models
leading to the time-dependent Gutzwiller approximation. 

 \subsection{Gutzwiller energy functionals for multi-band Hubbard models}
\label{sec2}
We first recall some results of the conventional Gutzwiller
approximation for the ground state properties of multi-band systems. 
We aim to study the physics 
of the following family of models, 
\begin{equation}\label{h2}
\hat{H}=\sum_{i\neq j} \sum_{\sigma,\sigma'}
t^{\sigma,\sigma'}_{i,j} \hcd_{i,\sigma}\hc_{j,\sigma'}+
\sum_i \hat{H}_{i,{\rm loc}}\;,
\end{equation}
where $t^{\sigma,\sigma'}_{i,j}$ denotes the `hopping parameters'  and 
 the operators $\hat{c}^{(\dagger)}_{i,\sigma}$ annihilate (create)
 an electron with spin-orbital index $\sigma$ on a lattice site $i$. 
The  local Hamiltonian 
\begin{eqnarray}\label{4.10a}
\hat{H}_{i;{\rm loc}}&=&\sum_{\sigma_1,\sigma_2}\varepsilon_{i;\sigma_1,\sigma_2}
\hcd_{i,\sigma_1} \hc_{i,\sigma_2}\\\nonumber
&&
+\sum_{\sigma_1,\sigma_2,\sigma_3,\sigma_4}
U_i^{\sigma_1,\sigma_2,\sigma_3,\sigma_4}
\hcd_{i,\sigma_1} \hcd_{i,\sigma_2}\hc_{i,\sigma_3} \hc_{i,\sigma_4}\;
\end{eqnarray}
is determined  by the orbital-dependent on-site energies  
$\varepsilon_{i;\sigma_1,\sigma_2}$
 and by the two-particle Coulomb interaction 
$U_i^{\sigma_1,\sigma_2,\sigma_3,\sigma_4}$.
Upon introducing the eigenstates $\ket{\Gamma}_i$ and eigenvalues 
$E_{i;\Gamma}$ of ~(\ref{4.10a}) 
 (which can be readily calculated by means of standard numerical techniques)
the local Hamiltonian can be written as
\begin{eqnarray}\label{eft}
\hat{H}_{i,{\rm loc}}&=&\sum_{\Gamma}
E_{i;\Gamma}|  \Gamma\rangind{i}
\langind{i} \Gamma'| \;.
\end{eqnarray}

Multi-band Gutzwiller wave-functions have the form
\begin{equation}\label{1.3}
|\Psi_{\rm G}\rangle=\hat{P}_{\rm G}|\Psi_{\rm S}\rangle=\prod_{i}\hat{P}_{i}|\Psi_{\rm S}\rangle\;,
\end{equation}
where $|\Psi_{\rm S}\rangle$ is a normalised Slater determinant 
and the local Gutzwiller correlator is defined as 
\begin{equation}
\hat{P}_{i}=\sum_{\Gamma,\Gamma^{\prime}}\lambda_{i;\Gamma,\Gamma^{\prime}}
|\Gamma \rangle_{i} {}_{i}\langle \Gamma^{\prime}|   \;.
 \end{equation}
Here we introduced the variational-parameter matrix 
$\lambda_{i;\Gamma,\Gamma^{\prime}}$ which allows us to optimise the occupation 
 and the form of the eigenstates  of $\hat{P}_{i}$.

The evaluation of expectations values with respect to the 
 wave function~(\ref{1.3}) is a difficult many-particle problem, 
which cannot be 
 solved in general.  As shown in~Refs.~\cite{buenemann1998,buenemann2005}, 
one can derive analytical expressions for the 
 variational ground-state energy in the limit of infinite spatial dimensions 
($D\to \infty$). Using this energy functional 
 for the study of finite-dimensional systems  is usually 
denoted as the `Gutzwiller approximation'. This approach is the basis 
of most applications of Gutzwiller wave functions in studies 
of real materials and it will also be used in this work. 
One should keep in mind, however, that the Gutzwiller approximation has its 
limitations and the study of some phenomena requires an evaluation 
 of expectation values in finite dimensions~\cite{buenemann2011d,buenemann2013}.

 For the evaluation of Gutzwiller wave functions in 
 infinite dimensions it is most convenient to impose the 
following (local) constraints~\cite{buenemann1998,buenemann2005} 
\begin{eqnarray}\label{1.10a}
\langle\hat{P}^{\dagger}\hat{P}^{}\rangle_{\Psi_{\rm S}}&=&1\;,\\
\label{1.10b}
\langle  \hat{P}^{\dagger}\hat{P}^{} \hat{c}^{\dagger}_{\sigma} \
\hat{c}^{}_{\sigma'}\rangle_{\Psi_{\rm S}}&=&\langle
 \hat{c}^{\dagger}_{\sigma}\hat{c}^{}_{\sigma'}   \rangle_{\Psi_{\rm S}}\;.~
\end{eqnarray}
With these constraints, the expectation value of the local 
Hamiltonian~(\ref{eft}) reads
\begin{equation}\label{kdr} 
\langle 
\hat{H}_{\rm loc}\rangle_{\Psi_{\rm G}}
=\sum_{\Gamma,\Gamma_1,\Gamma_2}E_{\Gamma}
\lambda_{\Gamma,\Gamma_1}^{*}\lambda_{\Gamma,\Gamma_2}^{}
m_{\Gamma_1,\Gamma_2}\;,
\end{equation}
where
\begin{equation}\label{dql}
m_{\Gamma_1,\Gamma_2}\equiv 
\langle(| \Gamma_1 \rangle   \langle \Gamma_2| )\rangle_{\Psi_{\rm S}}
\end{equation}
can be calculated by means of Wick's theorem.

The expectation value of a hopping operator in infinite dimensions 
has the form
\begin{equation}\label{8.410} 
\big \langle  \hat{c}_{i,\sigma_1}^{\dagger}\hat{c}_{j,\sigma_2}^{\phantom{+}} \big \rangle_{\Psi_{\rm G}}
=\sum_{\sigma'_1,\sigma'_2}q_{\sigma_1}^{\sigma'_1}\left( q_{\sigma_2}^{\sigma'_2}\right)^{*}\big \langle  
\hat{c}_{i,\sigma'_1}^{\dagger}\hat{c}_{j,\sigma'_2}^{\phantom{+}} \big \rangle_{\Psi_{\rm S}}\;,
\end{equation}
where the (local) renormalisation matrix $ q_{\sigma}^{\sigma'}$ is 
an analytic function
 of the variational parameters and of the non-interacting local density matrix
\begin{equation}
C_{i;\sigma,\sigma'}=\langle \hcd_{i,\sigma}\hc_{i,\sigma'}   \rangle_{\Psi_{\rm S}}\;.
\end{equation} 
The explicit form of the renormalisation matrix is given, e.g., 
in Ref.~\cite{buenemann2012} but will not be needed for our further
considerations in this work.~In the following, we assume that the correlated 
 and the non-correlated  local 
 density matrix are equal,
\begin{equation}
C^{\rm c}_{i;\sigma,\sigma'}=\langle \hcd_{i,\sigma}\hc_{i,\sigma'}   \rangle_{\Psi_{\rm G}}=C_{i;\sigma,\sigma'}\;.
\end{equation} 
This is the case when all non-degenerate orbitals on a lattice site
 belong to different representations of its point symmetry group. 

In summary, the expectation value of the Hamiltonian~(\ref{h2}),
\begin{equation}
E^{\rm GA}=E^{\rm GA}(\tilde{\lambda}^{(*)},\tilde{\rho})\;,
 \end{equation}
is a function 
  of all variational parameters 
$\tilde{\lambda}^{(*)}\equiv \{\lambda^{(*)}_{i;\Gamma,\Gamma'}\} $ and of the
 non-interacting density matrix $\tilde{\rho}$
with the elements 
\begin{equation}
\rho_{(i\sigma),(j\sigma')}\equiv
\langle \hat{c}_{j,\sigma'}^{\dagger}
\hat{c}_{i,\sigma}^{\phantom{+}}\rangle_{\Psi_{\rm S}}\;.
\end{equation} 
The same holds for the constraints~(\ref{1.10a}), (\ref{1.10b}),  for
 which we will use the abbreviation
 \begin{equation}\label{ko2}
g_n(\tilde{\lambda}^{(*)},\tilde{\rho}))=0\;\;\;,\;\;\;
1\leq n\leq n_{\rm c} 
\end{equation}
where $n_{\rm c}$ is the maximum number of independent constraints.
In Sec.~\ref{sec:oneb} we apply these results to
the special case of a single-band Hubbard model.

\subsection{Evaluation of time-dependent matrix elements}\label{secmat}
In this section, we will apply the concept, introduced in 
Sec.~\ref{sec:vp}, to the general class of Gutzwiller wave functions
\begin{equation}\label{1.3b}
|\Psi_{\rm G}(t)\rangle=\hat{P}_{\rm G}(t)|\Psi_{\rm S}(t)\rangle
=\prod_{i}\hat{P}_{i}(t)|\Psi_{\rm S}(t)\rangle\;,
\end{equation}
where the single-particle product states $\ket{\Psi_{\rm S}(t)}$ and the local
 correlation operators
\begin{equation}
\hat{P}_{i}(t)=\sum_{\Gamma,\Gamma^{\prime}}\lambda_{i;\Gamma,\Gamma^{\prime}}(t)
|\Gamma \rangle_{i} {}_{i}\langle \Gamma^{\prime}|  
 \end{equation}
are now time-dependent quantities.~

The state $|\Psi_{\rm S}(t)\rangle$ can be written as 
\begin{equation}\label{asc}
|\Psi_{\rm S}(t)\rangle=\prod_{\gamma}[\hat{h}^{\dagger}_{\gamma}(t)]^{n_{\gamma}}\ket{\rm vac}\;.
\end{equation}
Here, $n_{\gamma}\in (0,1)$ determines which of the single 
 particle states $\ket{\gamma(t)}$, described by the operators
\begin{equation}\label{567}
\hat{h}^{\dagger}_{\gamma}(t)=\sum_{\upsilon}u_{\upsilon,\gamma}(t)\hat{c}^{\dagger}_{\upsilon}
\;,
\end{equation}
are occupied and $u_{\upsilon,\gamma}(t)$ is a (time-dependent) unitary transformation,
\begin{equation}\label{9uh}
\sum_{\gamma}u_{\upsilon_1,\gamma}^*(t)u_{\upsilon_2,\gamma}(t)=\delta_{\upsilon_1,\upsilon_2}\;.
\end{equation}
The functions $u_{\upsilon,\gamma}$ 
constitute  a second set
of (time-dependent) variational parameters 
(in addition to $\lambda_{i;\Gamma,\Gamma^{\prime}}(t)$) 
and determine the single-particle
wave function $|\Psi_{\rm S}\rangle$. 
Note that the time dependence of the operators~(\ref{567}) implies that the 
non-interacting density matrix
\begin{equation}\label{rhou}
\rho_{\upsilon,\upsilon'}(t)=\langle \hat{c}^{\dagger}_{\upsilon'}\hat{c}^{}_{\upsilon}\rangle_{\Psi_{\rm S}(t)}=\sum_{\gamma}n_{\gamma}u^*_{\upsilon',\gamma}(t)u_{\upsilon,\gamma}(t)
\end{equation}
is also time dependent.

We start with a consideration of the time derivative in Eq.~(\ref{eq:lagran})
 which requires the evaluation of
\begin{equation}\label{qsf}
\frac{\langle \Psi_{\rm G} |\dot{\Psi}_{\rm G}\rangle}
{{\langle \Psi_{\rm G} |\Psi_{\rm G}\rangle}}
= 
\frac{\langle \Psi_{\rm S}| \hat{P}^{\dagger}_{\rm G} 
\hat{P}_{\rm G}|\dot\Psi_{\rm S}\rangle}
{\langle \Psi_{\rm S}| \hat{P}^{\dagger}_{\rm G} 
\hat{P}_{\rm G}|\Psi_{\rm S}\rangle}
+
\frac{
\langle \Psi_{\rm S}| \hat{P}^{\dagger}_{\rm G} 
\dot{\hat{P}}_{\rm G}|\Psi_{\rm S}\rangle}
{\langle \Psi_{\rm S}| \hat{P}^{\dagger}_{\rm G} 
\hat{P}_{\rm G}|\Psi_{\rm S}\rangle}
\end{equation}
and its complex conjugate.
With Eqs.~(\ref{asc}) and~(\ref{567}), we find
\begin{eqnarray}
|\dot\Psi_{\rm S}\rangle&=&\sum_{\gamma}\dot{\hat{h}}^{\dagger}_{\gamma}
\hat{h}^{}_{\gamma}|\Psi_{\rm S}\rangle\\
&=&\sum_{\upsilon}\sum_{\gamma,\gamma'}\dot{u}_{\upsilon,\gamma}u^*_{\upsilon,\gamma'}
\hat{h}^{\dagger}_{\gamma'}
\hat{h}^{}_{\gamma}|\Psi_{\rm S}\rangle\;.
\end{eqnarray}
This equation allows us to evaluate the contribution of the first term 
on the  r.h.s.~of Eq.~(\ref{qsf}) as
\begin{eqnarray}\nonumber
&&\frac{\langle \Psi_{\rm S}| \hat{P}^{\dagger}_{\rm G} \hat{P}_{\rm G}|\dot\Psi_{\rm S}\rangle}{\langle \Psi_{\rm S}| \hat{P}^{\dagger}_{\rm G} \hat{P}_{\rm G}|\Psi_{\rm S}\rangle}\\
\label{li123}
&&=\sum_{\upsilon}\sum_{\gamma,\gamma'}\dot{u}_{\upsilon,\gamma}u^*_{\upsilon,\gamma'}\frac{\langle \Psi_{\rm S}| \hat{P}^{\dagger}_{\rm G} \hat{P}_{\rm G}
\hat{h}^{\dagger}_{\gamma'}
\hat{h}^{}_{\gamma}|\Psi_{\rm S}\rangle}
{\langle \Psi_{\rm S}| \hat{P}^{\dagger}_{\rm G} \hat{P}_{\rm G}|\Psi_{\rm S}\rangle}\\
&&=\sum_{\upsilon,\gamma}n_{\gamma}\dot{u}_{\upsilon,\gamma}u^*_{\upsilon,\gamma}\;.
\end{eqnarray}
In the last line, we have used that, in all relevant applications,
 $\hat{P}_{\rm G}$ and $|\Psi_{\rm S}\rangle$ have the same symmetry and, therefore,
 all contributions with $\gamma\ne \gamma'$ vanish in~(\ref{li123}). 

We now proceed with a consideration of the second term on the 
r.h.s.~of Eq.~(\ref{qsf}).~With the
 definition of the correlation operator $\hat{P}_{G}$ we find
\begin{equation} \label{ghj}
\langle \Psi_{\rm S}| \hat{P}^{\dagger}_{\rm G} 
\dot{\hat{P}}_{\rm G}|\Psi_{\rm S}\rangle=
\sum_i\Big \langle \Psi_{\rm S}\Big |\Big (\prod_{j(\ne i)}\hat{P}^{\dagger}_{j} 
\hat{P}_{j}\Big ) 
\hat{P}^{\dagger}_{i}\dot{\hat{P}}_{i}\Big  |\Psi_{\rm S}\Big \rangle\;.
\end{equation}
The r.h.s.~of~(\ref{ghj}) can be evaluated by the standard 
 diagrammatic techniques in infinite dimensions~\cite{buenemann1998}.~This leads to
\begin{eqnarray} \nonumber
&&\frac{\langle \Psi_{\rm S}| \hat{P}^{\dagger}_{\rm G} 
\dot{\hat{P}}_{\rm G}|\Psi_{\rm S}\rangle}
{\langle \Psi_{\rm S}| \hat{P}^{\dagger}_{\rm G} 
\hat{P}_{\rm G}|\Psi_{\rm S}\rangle }=
\sum_i\langle \Psi_{\rm S}| \hat{P}^{\dagger}_{i} 
\dot{\hat{P}}_{i}|\Psi_{\rm S}\rangle\\\label{uis}
&&\,\,\,\,\,\,=\sum_i\sum_{\Gamma_1,\Gamma_2,\Gamma_3}
\lambda^*_{i;\Gamma_1,\Gamma_2}\dot{\lambda}_{i;\Gamma_1,\Gamma_3}
m_{i;\Gamma_2,\Gamma_3}\;,
\end{eqnarray}
where $m_{i;\Gamma_2,\Gamma_3}=m_{i;\Gamma_2,\Gamma_3}(t)$, 
as defined in Eq.~(\ref{dql}), depends on the local elements of the density
 matrix~(\ref{rhou}). 

\subsection{Lagrangian and equations of motion}\label{df67c}
From Eqs.~(\ref{li123}),(\ref{uis}), together with the 
expectation value of the Gutzwiller energy derived in Sec.~\ref{sec2},
we are now in the position to derive the Lagrangian Eq.~(\ref{eq:lagran}).
However, we also need to include two sets of constraints, 
(i) the unitarity of  $u^{}_{\upsilon,\gamma}$ and (ii) the Gutzwiller 
 constraints~(\ref{ko2}).
Therefore we finally obtain the following Lagrangian
\begin{eqnarray}
L &=&\frac{\rmi}{2}\sum_i\sum_{\Gamma_1,\Gamma_2,\Gamma_3}\left\lbrack
\lambda^*_{i;\Gamma_1,\Gamma_2}\dot{\lambda}_{i;\Gamma_1,\Gamma_3}
- \dot{\lambda}^*_{i;\Gamma_1,\Gamma_2} {\lambda}_{i;\Gamma_1,\Gamma_3}
\right\rbrack m_{i;\Gamma_2,\Gamma_3}  \label{lagfin} \\
&+& \frac{\rmi}{2}\sum_{\upsilon,\gamma}n_{\gamma}\left\lbrack
\dot{u}_{\upsilon,\gamma}u^*_{\upsilon,\gamma}
- {u}_{\upsilon,\gamma}\dot{u}^*_{\upsilon,\gamma} \right\rbrack -
E^{\rm GA}(\tilde{\lambda}^{(*)},\tilde{\rho}) \nonumber \\
&-& \sum_{\upsilon,\gamma,\gamma'}\Omega_{\gamma,\gamma'}(t)
(u^*_{\upsilon,\gamma}u_{\upsilon,\gamma'}-1)
- \sum_n \Lambda_n(t)g_n({\lambda}^{(*)},u^{(*)}_{\upsilon,\gamma})\nonumber
\end{eqnarray}
where  $\Lambda_n(t)$ and $\Omega_{\gamma,\gamma'}(t)$ are
(real) Lagrange-parameters.
As will be exemplified below in the single-band case, the original
Hamiltonian can be time dependent which will reflect in a
time dependence of $E^{\rm GA}$ and allows for a 
coupling with arbitrary external fields. 


From Eq.~(\ref{lagfin}) the Euler-Lagrange equations can now be
derived in the standard way.
The equation for 
the variational parameters $u^{*}_{\upsilon,\gamma}$
 reads
 \begin{equation}\label{xcu}
-\rmi \dot{u}_{\upsilon,\gamma}
+\frac{\partial }{\partial u^{*}_{\upsilon,\gamma}}
\left(E^{\rm GA} +V^{\rm GA}_t+V^{\lambda}+\sum_n\Lambda_ng_n\right)
+\sum_{\upsilon_2}\Omega_{\upsilon,\upsilon_2}
u_{\upsilon_2,\gamma}=0
\end{equation}
which in terms of the density matrix~(\ref{rhou}) can be
rewritten as
\begin{equation}\label{eq1}
\rmi \dot{\tilde{\rho}}=[\tilde{h}^{\rm GA},\tilde{\rho}]\;.
\end{equation}
Here we have introduced the `Gutzwiller Hamiltonian'
\begin{equation}
h^{\rm GA}_{\upsilon,\upsilon'}\equiv 
\frac{\partial}{\partial \rho_{\upsilon',\upsilon}}
\bigg( E^{\rm GA}+V^{\lambda}+\sum_n\Lambda_n g_n\bigg)
\end{equation}
and a potential $V^{\lambda}$ which depends on the (time-dependent) 
phases of $\lambda_{\Gamma,\Gamma'}$
\begin{equation}
V^{\lambda}=\frac{\rmi}{2}\sum_i\sum_{\Gamma_1,\Gamma_2,\Gamma_3}\left\lbrack
\lambda^*_{i;\Gamma_1,\Gamma_2}\dot{\lambda}_{i;\Gamma_1,\Gamma_3}
- \dot{\lambda}^*_{i;\Gamma_1,\Gamma_2} {\lambda}_{i;\Gamma_1,\Gamma_3}
\right\rbrack m_{i;\Gamma_2,\Gamma_3}\,.
\end{equation}

Note that the same equation of motion for $\tilde{\rho}$ is obtained in 
the previous formulation
of the TDGA \cite{goe01,goe03,goe042,goe08}.
The new ingredient in the present formulation is the implementation
of the explicit time dependence of the variational parameters 
$\lambda^{*}_{i;\Gamma_1,\Gamma_2}$.~It is obtained from Eq.~(\ref{lagfin}) as
 \begin{eqnarray}\label{eq2}
&&\rmi \sum_{\Gamma_3}\left(\dot{\lambda}_{i;\Gamma_1,\Gamma_3}
m_{i;\Gamma_2,\Gamma_3}+\frac{1}{2}{\lambda}_{i;\Gamma_1,\Gamma_3}
\dot{m}_{i;\Gamma_2,\Gamma_3}\right)
\\\nonumber
&&= \frac{\partial}
{ \partial  \lambda^*_{i;\Gamma_1,\Gamma_2}}
\left(
 E^{\rm GA}({\lambda}^{(*)},\tilde{\rho})
+\sum_n\Lambda_ng_n({\lambda}^{(*)},\tilde{\rho})
\right)\;.
 \end{eqnarray}
Equations~(\ref{eq1}) and~(\ref{eq2}) for $\tilde{\rho}(t)$ and 
$\lambda_{i;\Gamma,\Gamma'}(t)$ constitute the time-dependent 
Gutzwiller theory for multi-band Hubbard models.~

\section{Example: The one-band model}\label{sec:oneb}
\subsection{Evaluation of the time-dependent GA energy} 
In order to make the results, derived in the previous section, more 
transparent we consider the case of
the single band Hubbard model,
\begin{equation}\label{one_band}
\hat{H}(t)=\sum_{i,j}\sum_{\sigma=\uparrow,\downarrow}t_{i,j}(t)
\hat{c}_{i,\sigma}^{\dagger}\hat{c}_{j,\sigma}^{\phantom{+}}+\sum_{i}\hat{H}_{i,{\rm loc}}(t)\;,
\end{equation}
where $t_{i,j}$ denotes the `hopping parameters' 
($t_{i,i}\equiv 0$) and the operators 
$\hat{c}^{(\dagger)}_{i,\sigma}$ annihilate (create) an electron with 
spin index $\sigma$ on a lattice site $i$. We further introduced 
\begin{equation}\label{one_bandloc}
\hat{H}_{i,{\rm loc}}(t)=\sum_{\sigma} v_{i,\sigma}(t) \hat{n}_{i,\sigma} +U_i(t) \hat{n}_{i,\uparrow}\hat{n}_{i,\downarrow}
\end{equation}
and $\hat{n}_{i,\sigma} \equiv \hat{c}^{\dagger}_{i,\sigma}\hat{c}_{i,\sigma}$.
All parameters in the Hamiltonian can be time and
spatial-dependent
 which allows us to study the response to scalar fields $v_i(t)$, vector
potential fields through the Peierls substitution, 
\begin{eqnarray}
  \label{eq:peierls}
&&t_{i,j}\rightarrow t_{i,j} e^{\rmi  \phi_{i,j}}\;,  \nonumber\\
&&\phi_{i,j}=-q\int_{\vec{R}_i}^{\vec{R}_j}\vec{A}\cdot d\vec{ r}\;,
\end{eqnarray}
and modulations of the on-site interaction. 
Here, we introduced $q=-|e|$ for electrons. 

The local Hamiltonian can be diagonalised by the states 
$\ket{\Gamma}_i=\ket{d}_i,\ket{\sigma}_i,\ket{\emptyset}_i$
in which the site $i$ is double
occupied (i.e. has a doublon), single occupied by an electron with spin $\sigma$, or
empty. 
Restricting for simplicity to  paramagnetic states, we can work with a 
diagonal matrix of variational parameters,
$\lambda_{i\Gamma\Gamma^{\prime}}=\lambda_{i\Gamma}\delta_{\Gamma\Gamma^{\prime}}.$
Thus the local `Gutzwiller projection operator' reads, 
\begin{equation}\label{pi_one}
 \hat{P}_{i}=\lambda_{d,i}|d\rangle_i {}_{i}\langle d|+
\lambda_{i,\uparrow}|\!\!\uparrow\rangle_i {}_{i}\langle\uparrow\! |+
\lambda_{i,\downarrow}\!|\!\downarrow\rangle_i {}_{i}\langle\downarrow\! |
+\lambda_{\emptyset,i}|\emptyset\rangle_i {}_{i}\langle\emptyset |\;,
\end{equation}
where the variational parameters $\lambda_{i;\Gamma}$
are related to the probability $p_{i,\Gamma}$ of finding 
a configuration $\Gamma$ at site $i$ according to
\begin{equation}
  \label{eq:probgamma}
p_{i\Gamma}\equiv \langle {\Psi_{\rm
    G}}|  \Gamma\rangind{i}
\langind{i} \Gamma|{\Psi_{\rm
    G}}\rangle =|\lambda_{i\Gamma}|^2 m_{i\Gamma}\;.
\end{equation}

We have four configuration probabilities per site which we denote as
$p_{i,\Gamma}\equiv E_i$, $S_{i,\sigma}$, $D_i$ for empty, single, and
doublon occupied sites.
In the present case, the constraints (\ref{1.10a}),(\ref{1.10b}) read,
\begin{eqnarray}
D+S_{\uparrow}+S_{\downarrow}+E\Big|_i&=&1\;, \label{conprob}\\
D+S_{\sigma}&=&\rho_{\sigma}\Big|_i \label{con2prob}\;,
\end{eqnarray}  
where $...\Big|_i$ indicates that the index $i$ is implicit
everywhere in the expression.  The first constraint is the statement 
 $\sum_{\Gamma}p_\Gamma=1$, as it should be, while the second constraint
 implies that  local charges are the same in the 
correlated and the uncorrelated state. 
Obviously this also guarantees that the total charge per spin in the system is
the same in both states.
We will show below that these constrains lead to
equations of motions that nicely respect charge conservation. 

In the real-space basis the index $\upsilon$ in Eq.~(\ref{rhou})
stands for $i,\sigma$. For paramagnetic states the uncorrelated density matrix
is diagonal with respect to the spin variables,
$\rho_{i,\sigma;j,\sigma'}\equiv\rho_{i,j;\sigma}\delta_{\sigma,\sigma'}$. We
will also use the short-hand notation, 
$\rho_{i\sigma}\equiv \rho_{i\sigma,i\sigma} $. 

According to Eq.~(\ref{8.410})  the expectation value of the one-band
 hopping operator in infinite dimensions can be written as
\begin{equation}
\big \langle
\hat{c}_{i,\sigma}^{\dagger}\hat{c}_{j,\sigma}^{\phantom{+}} \big
\rangle_{\Psi_{\rm G}} 
=q_{i,\sigma}q_{j,\sigma}^{*} \rho_{j,i,\sigma} \;,
\end{equation}
where the (local) renormalisation factors  are given by
\begin{equation}
q_{\sigma}=\lambda^*_{\sigma} \lambda_{\emptyset}
 (1-\rho_{\bar{\sigma}})+\lambda^*_{d}\lambda_{\bar{\sigma}}
\rho_{\bar{\sigma}}\Big|_i\label{eq:renorm1b}
\end{equation} 
and we used the notation
\begin{equation}
\bar{\uparrow}=\downarrow\;\;\; {\rm and}\;\;\;  \bar{\downarrow}=\uparrow\;.
\end{equation}
The $q_{\sigma}$ factors renormalise the probability amplitude for the
annihilation of an electron on site $j$ and the creation of an electron on site
$i$. Each one of these processes
has two possible channels. For example the creation of an
electron at $i$ with spin up can be seen as a transition from an empty
state  to an $\ket{\uparrow}$  state [leading to the first term in
Eq.~(\ref{eq:renorm1b})] or a transition from $\ket{\downarrow}$  
to a doublon state [leading to second term  in
Eq.~(\ref{eq:renorm1b})].  Since the variational parameters are now
complex,  these two channels can
interfere either constructively or destructively affecting the total
hopping amplitude. This issue
is discussed further in \ref{ap:phase} where the physical origin of this
renormalisation is exemplified in a simple two-site case. 

It is convenient to write the parameters $ \lambda_\Gamma$ in terms of a real positive
amplitude and a phase
\begin{equation}
  \label{eq:amplphas}
  \lambda_\Gamma=\sqrt{\frac{p_\Gamma}{m_\Gamma}}e^{\rmi\varphi_\Gamma}\Big|_i\;,
\end{equation}
which are used in the following as the dynamical 
 variables. With these definitions the hopping renormalisation factors
read, 
\begin{equation}
q_{\sigma}=e^{-\rmi\chi_\sigma}(q_{\emptyset,\sigma}+q_{d,\sigma}
e^{-\rmi\eta})\Big|_i\label{eq:renormeta} 
\end{equation} 
with the definitions for site $i$,
\begin{eqnarray}
  q_{\emptyset,\sigma}&\equiv &\sqrt{\frac{S_\sigma E}{m_\sigma
      m_\emptyset}}(1-\rho_{\bar{\sigma}})=\left.\sqrt{\frac{(1-\rho_\uparrow-\rho_\downarrow+D)(\rho_\sigma-D)}{\rho_\sigma
    (1-\rho_\sigma)}} \;\right|_i\;, \label{eq:qempty}\\
 q_{d,\sigma}&\equiv &\sqrt{\frac{D S_{\bar \sigma} }{m_d m_{\bar \sigma}}} \rho_{\bar{\sigma}}\;,=\left. \sqrt{\frac{D(\rho_{\bar\sigma}-D)}{\rho_\sigma
   (1-\rho_{\sigma})}}\;\right|_i\;,\label{eq:qd}\\
\eta&\equiv&\varphi_{d}+\varphi_{\emptyset}-\varphi_{\uparrow}-\varphi_{\downarrow}\Big|_i\;,\\
\chi_\sigma&\equiv&\varphi_{\sigma}-\varphi_{\emptyset}\Big|_i\;.
\end{eqnarray}
The phases   $\varphi_{i,\sigma}$,
$\varphi_{d,i}$ have been eliminated in favor of $\chi_{i,\sigma}$ and
$\eta_{i,\sigma}$. Note that $\varphi_{\emptyset,i}$ does not appear
anywhere in the functional and therefore can be disregarded as
a dynamical variable. 
In addition, we have 
used the constraints Eqs.~(\ref{conprob}),(\ref{con2prob}) to eliminate $E$ and $S_\sigma$
in favor of $D$. Therefore our dynamical variables are the single-particle 
amplitudes $u_{v,\gamma}$, the double occupancy $D_i$ and the phases $\chi_{i,\sigma}$ and
$\eta_{i}$. 

In summary, the expectation value of
the time-dependent single-band Hubbard model,
\begin{equation}
\frac{\langle \Psi_G(t)|\hat{H}(t) | \Psi_G(t)
   \rangle} 
{ \langle \Psi_G(t) | \Psi_G(t)  \rangle}=E^{\rm GA}(\tilde{\rho},D_i,\eta_{i},\chi_{i,\sigma})\;,
 \end{equation}
is a function 
  of the variational parameters and of the
 non-interacting density matrix $\tilde{\rho}$,
 \begin{equation}
   \label{eq:egahub}
   E^{\rm GA}=\sum_{i,j\sigma}t_{i,j}q_{i,\sigma}q_{j,\sigma}^{*}
   \rho_{j,i,\sigma} +
\sum_i U_i D_i+\sum_{i\sigma}v_{i\sigma}\rho_{i\sigma}\;.
 \end{equation}

\subsection{Lagrangian and equations of motion}\label{df67c1b}
We are now in the position to evaluate the Lagrangian Eq.~(\ref{lagfin})
which can be written
as 
  \begin{eqnarray}
  \label{eq:lagran1b}
L&=&-\sum_i D_i(U_i+\dot \eta_i) - \sum_{i,\sigma} \rho_{i,\sigma}
 (v_{i,\sigma}+ \dot\chi_{i,\sigma})\nonumber\\
&-&\sum_{i,j\sigma=\uparrow,\downarrow}t_{i,j} 
e^{\rmi (\chi_{j,\sigma}-\chi_{i,\sigma})}(q_{\emptyset,j,\sigma}+q_{d,j,\sigma}e^{\rmi \eta_j})
  (q_{\emptyset,i,\sigma}+q_{d,i,\sigma} e^{-\rmi \eta_i})\rho_{j,i;\sigma}\\
&+&\rmi \sum_{\upsilon,\gamma}n_{\gamma}\dot{u}_{\upsilon,\gamma}u^*_{\upsilon,\gamma}-
\sum_{\upsilon,\upsilon'}\Omega_{\upsilon,\upsilon'}(t)
\left(\sum_{\gamma}u^*_{\upsilon,\gamma}u_{\upsilon',\gamma}-1\right)
    \end{eqnarray}
Note that, since we have implemented the constraints 
(\ref{ko2}) explicitly, the corresponding Lagrange-parameter terms 
 are not needed.  

The Lagrangian is invariant with respect to a gauge transformation of 
the form, 
\begin{eqnarray*}
  \label{eq:gauge}
  u_{i,\sigma,\upsilon}(t)&\rightarrow& u_{i,\sigma,\upsilon}(t) e^{-\rmi 
\chi_{i\sigma}'(t)}\\
\chi_{i,\sigma}(t)&\rightarrow&\chi_{i,\sigma}(t)+\chi_{i,\sigma}'(t). 
\end{eqnarray*}
 Notice that the hopping amplitude and the site energy 
 transform in a way that generalises to the lattice the  usual gauge
transformation in the continuum ($ q {\bf A}\rightarrow  q {\bf
  A}-\nabla  \chi$,  $v\rightarrow v+\dot \chi$). Hence, 
$\chi_{i,\sigma}$ plays the role of a gauge phase and implements  
charge conservation. Indeed, the Euler-Lagrange equations for
  $\chi_{i,\sigma}$ yield the usual charge conservation law, 
\begin{equation}
  \label{eq:contin}
  \dot  \rho_{i,\sigma} =\sum_j j_{i,j}\;,
\end{equation}
where the current in a bond is given by 
\begin{equation}
  \label{eq:cur}
  j_{i,j}=\rmi [t_{i,j} e^{\rmi(\chi_{j,\sigma}-\chi_{i,\sigma})} (q_{\emptyset,j,\sigma}+q_{d,j,\sigma}e^{\rmi\eta_j})
  (q_{\emptyset,i,\sigma}+q_{d,i,\sigma} e^{-\rmi\eta_i})\rho_{j,i;\sigma}-{\rm h.c.}]\;.
\end{equation}

The Euler-Lagrange equations for the  variational parameters
$u^{*}_{\upsilon,\gamma}$  yield again the equation of motion
for the density matrix
\begin{equation}\label{78e}
\rmi \dot{\tilde{\rho}}=[\tilde{h}^{\rm GA},\tilde{\rho}]\;,
\end{equation}
with the `Gutzwiller Hamiltonian'
 matrix 
\begin{equation}
h^{\rm GA}_{\upsilon,\upsilon'}\equiv 
\frac{\partial}{\partial \rho_{\upsilon',\upsilon}} \left( E^{\rm
    GA}+\sum_{i,\sigma} \rho_{i,\sigma} \dot\chi_{i,\sigma}\right) 
\end{equation}
and the last term ensuring gauge invariance. 
 
In addition to (\ref{78e}), one obtains equations of motion for the 
double-occupancy parameters and the phases.  
For $\eta_i$ we obtain the Euler-Lagrange equation 
\begin{equation}
  \label{eq:d}
  \dot D_i=\rmi \sum_{j\sigma} [t_{i,j} e^{\rmi(\chi_{j,\sigma}-\chi_{i,\sigma})} 
(q_{\emptyset,j,\sigma}+q_{d,j,\sigma}e^{\rmi\eta_j})
  q_{d,i,\sigma} e^{-\rmi\eta_i}
   \langle  
\hat{c}_{i,\sigma}^{\dagger}\hat{c}_{j,\sigma}^{\phantom{+}} \big \rangle_{\Psi_{\rm S}}-{\rm h.c.}].\
\end{equation}
From Eq.~(\ref{eq:lagran1b}) we see that $\eta$ plays for $D$ a
similar role as the gauge phase $\chi$ for the charge. A time
dependent $\eta$ is  equivalent to a change in the Hubbard $U$.  
However, there are important differences. 
In the case of a uniform system for $n>1$ and $U\rightarrow \infty$ 
the probability to find empty sites $E \rightarrow 0$  
which leads to $q_{\emptyset,j,\sigma}\rightarrow 0$
[c.f., Eq.~(\ref{eq:qempty})]. Then $\eta$ becomes a gauge phase and 
$D_i$ is conserved as can be easily checked from the charge constraints. 
Using $E$ instead of $D$ as variational parameter one arrives to 
the conclusion that $E$ is conserved for  $n<1$ and $U\rightarrow
\infty$. In general, however, $\eta$ is not a gauge phase and $D$ is of
course not conserved. This reflects the fact that, when an electron
jumps from a doubly-occupied site $i$ to site $j$, one may have the
process $|d \rangle_i |\sigma \rangle_j  \rightarrow  |\sigma
\rangle_i |d \rangle_j $ which conserves $D$ but one can also
have the process $ |d \rangle_i |\emptyset \rangle_j  \rightarrow  |\sigma
\rangle_i |\bar\sigma \rangle_j $ which does not conserve
$D$. Therefore, in general, $\eta_i(t)$ is not arbitrary and has
observable physical consequences as will be explained for the 2-site example 
(cf. the following section). On the other hand since
$\chi_i(t)$ is a gauge phase we can work in a gauge in which $\chi_i(t)=0$. 
Finally form requiring stationarity with respect to $D$ we get
\begin{eqnarray}
  \label{eq:doteta}
  \dot\eta_i=-U_i&-&\left(\frac{\partial q^{\emptyset}_{i\sigma}}{\partial D_i} +\frac{\partial q^{d}_{i\sigma}}{\partial D_i} e^{-\rmi \eta_i}\right)\sum_{j\sigma}t_{i,j} 
e^{\rmi (\chi_{j\sigma}-\chi_{i\sigma})}(q_{\emptyset,j,\sigma}+q_{d,j,\sigma}e^{\rmi \eta_j})
  \rho_{j,i\sigma}\nonumber\\
&+&{\rm h.c.}
\end{eqnarray}


\subsection{Non interacting limit}
As a check of the consistency of the equations of motion it is
instructive to see how the non interacting limit is recovered when
$U_i\rightarrow 0$. In this case 
the double occupancy should factorise as $D_i=\rho_{i\uparrow}
\rho_{i\downarrow}$. Using this as an ansatz together with $\eta_i=0$ 
it is easy to check that Eq.~(\ref{eq:doteta}) is satisfied. This
follows from the fact that $q_{i,{\emptyset},\sigma}+q_{d,i,\sigma}$
is the hopping renormalisation of the static theory and attains its
maximum value $q_{\emptyset,i,\sigma}+q_{d,i,\sigma}=1$  as a
function of  $D_i$ precisely when  $D_i=\rho_{i\uparrow}
\rho_{i\downarrow}$. Thus its derivative as a function of $D_i$
evaluated at $D_i=\rho_{i\uparrow}
\rho_{i\downarrow}$
vanishes (as can be checked from an explicit computation)  and
Eq.~(\ref{eq:doteta}) is satisfied.   

Using the same ansatz we notice that in this limit 
Eq.~(\ref{eq:d}) can be written as, 
\begin{equation}
  \label{eq:dnon}
  \dot D_i=  \sum_{j\sigma} q_{d,i,\sigma} j_{i,j;\sigma}= \sum_{\sigma}\rho_{i,\bar\sigma}\dot\rho_{i,\sigma}
\end{equation}
which completes the consistency check. On the last passage, we used
Eqs.~(\ref{eq:qd}) and (\ref{eq:contin}).  For small $U$, one would
recover the time dependent Hartree-Fock approximation which in the
small amplitude limit corresponds to the usual RPA.

\subsection{Two-site example}
In order to clarify the meaning of the TDGA equations it is
interesting to consider the following two-site, two-electron example
whose exact time-dependent evolution can be found analytically. 
The Hamiltonian is assumed to be time-independent
with parameters $v_{i\sigma}=0$, $t_{1,2}=-t_0$; the interaction on
site 2 is infinite, $U_2=\infty$, while  $U_1$ is a  variable. 
Albeit simple, the problem is in the strong coupling
limit and provides a non-trivial test for the performance of the
theory. Even more, been a zero dimensional problem we expect it to be
the more demanding test bed for the TDGA which is based on infinite
dimension results. 

\subsubsection{Exact solution:}

The two-site Hamiltonian defined above is diagonalised by the states,
$$|\Psi_{\pm}\rangle=a_{\pm}|d\emptyset\rangle+b_{\pm}|s \rangle$$
with
$$|s \rangle\equiv\frac1{\sqrt2}(\hat{c}_{1\uparrow}^{\dagger}\hat{c}_{2\downarrow}^{\dagger}+\hat{c}_{2\uparrow}^{\dagger}\hat{c}_{1\downarrow}^{\dagger})|{\rm
vac}\rangle$$
and eigenvalues $E_{\pm}$ given by, 
\begin{eqnarray}
  \label{eq:eigenval2}
  E_{\pm}&=&\frac12(U\pm\omega_0)\;,\\
\omega_0&=&\sqrt{U^2+8t_0^2} \;.
\end{eqnarray}
The time-dependent wave function can be expanded as
$$
|\Psi(t)\rangle=\alpha_+ |\Psi_{+}\rangle e^{-\rmi E_+t}+\alpha_- |\Psi_-\rangle e^{-\rmi E_-t}\;,
$$
such that the double occupancy is given by 
$$D_1(t)= \langle\Psi(t)|d\emptyset \rangle \langle d\emptyset |\Psi(t)\rangle\;,
$$
where
$$
\langle d\emptyset |\Psi(t)\rangle= \alpha_+ a_+  e^{-\rmi  E_+t}+
\alpha_- a_-  e^{-\rmi  E_-t}\;.
$$
Independently of the initial condition, as long as there is a finite overlap
with both eigenstates,  the double occupancy has a
fluctuating part going like $\sim\cos(\omega_0t)$. 

As an example we choose the starting state at $t=0$ as
$$|\Psi(0)\rangle=\hat{c}_{1\uparrow}^{\dagger}\hat{c}_{1\downarrow}^{\dagger}|{\rm
vac}\rangle\equiv |d\emptyset\rangle\;. $$ 
The probability of finding the system in the state $|d\emptyset\rangle$
is then given by 
$$
D_1(t)=1-\frac{4t_0^2}{\omega_0^2}[1-\cos(\omega_0t)]\sim1-\frac{4t_0^2}{U^2}[1-\cos(\omega_0t)]\;,
$$
where the last approximate equality is valid for $U>>t_0$. Clearly the probability
to find the system in the state $|s \rangle$ is $1-D_1$ from which one finds,
$$n_{1}(t)=1+D_1(t)\;,$$
$$n_{2}(t)=1-D_1(t)\;,$$
with $n_i=n_{i\uparrow}+n_{i\downarrow}$.

\begin{figure}[tb]
$$\includegraphics[width=9cm,clip=true]{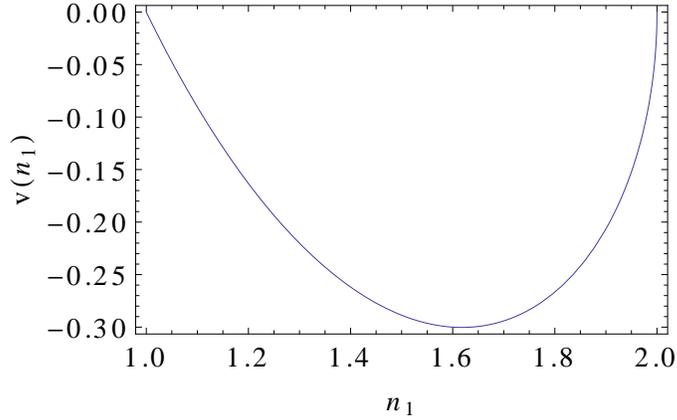}$$
\caption{Effective potential for charge fluctuations in the two site case.}     
\label{fig:vdn1}                                                   
\end{figure}                

\subsubsection{Time-dependent Gutzwiller approximation:}

To solve the TDGA equations it is useful to note that $U_2=\infty$
leads to $D_2(t)\rightarrow 0$. If the constraints where exactly
satisfied in the GA this would also imply $E_1(t)=0$. However, the
 numerical solution of the static GA shows that
$E_1$ is nonzero but very small for any $U_1$ ($E_1<0.08$) and
vanishes for  $U_1\to -\infty$. In the following we assume for simplicity
$E_1=0$. Then the constraint Eq.~(\ref{conprob}) implies  $D_1=n_1-1$
and one can evaluate the TDGA equations analytically.

The hopping renormalisation factors simplify to,
\begin{eqnarray*}
q_{1\sigma}&=&e^{-\rmi(\chi_{1\sigma}+\eta_1)}q_{d,1\sigma}\;,\\
q_{2\sigma}&=&e^{-\rmi\chi_{2\sigma}}q_{\emptyset,2\sigma}\;,
\end{eqnarray*} 
and the energy becomes
$$E^{GA}=-4 t_0
e^{-\rmi(\chi_{1\sigma}-\chi_{2\sigma}+\eta_1)}\left(1-\frac1n_1\right)
\rho_{21\sigma}+h.c.+U_1 (n_1-1)\;.$$
The bonding single-particle state is defined by
$$ \hat h_{1\sigma}^\dagger=u_{1\sigma} \hat c_{1\sigma}^\dagger +u_{2\sigma}
 \hat c_{2\sigma}^\dagger  \;.$$
Without loss of generality we set $u_1=\sqrt{\frac{n_1}2} e^{\rmi\phi}$ and 
$u_2=\sqrt{\frac{n_2}2}$ which yields 
$$\rho_{21\sigma}= e^{-\rmi\phi} \frac{\sqrt{n_1(2-n_1)}}2\;,$$
and
$$\dot u_1 u_1^*=\frac{\dot n_1}4+i\frac{n_1}2\dot \phi\;.$$
The Lagrangian reads, 
$$L=-(U_1+\dot \eta_1+\dot\phi)n_1 -4 t_0
\cos(\eta_1+\phi)v(n_1)\;,$$
where we chose a gauge in which $\chi=0$ and 
\begin{equation}\label{eq:poteff}
v(n_1)=-\left(1-\frac1n_1\right)
\sqrt{n_1(2-n_1)}\;.
\end{equation} 
Furthermore, since $\phi$ 
and $\eta_1$ play the same role, we can set $\phi=0$. Thus we get a
Lagrangian with two dynamical variables $n_1$ and $\eta_1$ which are
conjugate.  
Their equations of motion have the form,
\begin{eqnarray}
  \label{eq:eom2site}
  \dot n_1&=&-4 t_0 \sin(\eta_1)v(n_1)\;,\\ \label{eq:eom2siteb}
\dot \eta_1&=&-U_1-4 t_0 \cos(\eta_1)v'(n_1)\;.
\end{eqnarray}
The `potential' $v(n_1)$ is shown in Fig.~\ref{fig:vdn1}. 
The static solution is given by $\eta_1=0$ and $v'(n_1)=-U_1/(4t_0)$.
 Hence, for $U_1=0$  the static charge is $n_1^0=\frac{1}{2}
\left(1+\sqrt{5}\right)\approx 1.62$ and it decreases towards $n_1=1$ when
$U_1\rightarrow U_1^c=4t_0$ where a spurious Brinkman-Rice transition
occurs. For negative $U_1$ instead the charge tends asymptotically to
2 when  $U_1\rightarrow -\infty$. 

\begin{figure}[tb]
$$\includegraphics[width=9cm,clip=true]{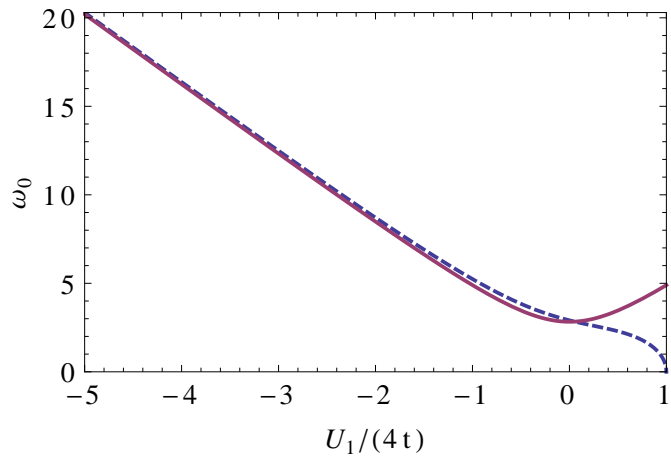}$$
\caption{Oscillatory frequency in the TDGA for small amplitudes
  (dashed line) and in the exact solution for arbitrary fluctuations (full line) .}     
\label{fig:w0du}                                                   
\end{figure}

Equations~(\ref{eq:eom2site}),(\ref{eq:eom2siteb})  
can be readily solved for small oscillations around the
static solution. We find that the oscillatory frequency is 
$\omega_0=4t_0\sqrt{v''(n_1^0) (-v(n_1^0))}$. For negative $U_1$ and
until $U_1=0$ is approached we find that this frequency is in
excellent agreement with the exact oscillation frequency of the
two-site problem (c.f., Fig.~\ref{fig:w0du}).   For positive $U_1$ as
the spurious Brinkman-Rice point is approached, not surprisingly
the approximations fail:  the exact frequency increases
monotonously as a function of $U_1$ while the approximate one vanishes
at the Brinkman-Rice point. We attribute this to the failure of the GA
in this finite-site system. Indeed we will show below that for
high-dimensions a similar softening approaching a Mott state is real. 
The softening can be traced back to the fact that in the Brinkman-Rice 
phase the doublon charge gets frozen at zero therefore it is conserved
and  $\eta_1$ becomes a gauge phase which leads to an energy independent of $\eta_1$. 
We will see that a similar phenomena is found in the usual
Brinkman-Rice transition.

\section{Linear response: The small amplitude limit}\label{sec:sma}
In the previous two-site model, the expansion of the effective potential 
$v(n_1)$, Eq.~(\ref{eq:poteff}), around the stationary value $n_1^0$
yields the dynamics in the vicinity of the GA saddle-point.
Such an expansion is also important for the evaluation of
response functions, where a (weak) external perturbation drives
the system out of equilibrium. 

Based on our general scheme derived in Sec.~\ref{df67c}, the small amplitude
limit is obtained by expanding the equations of motion,
Eqs.~(\ref{eq1}),(\ref{eq2}), around the ground state values 
$\tilde{\rho}^0$ and 
$\lambda^{0}_{\Gamma,\Gamma'}$,
\begin{eqnarray}
\tilde{\rho}(t)&\approx& \tilde{\rho}^0+\delta{{\rho}(t)}\;,\\
\lambda_{i;\Gamma,\Gamma'}(t)&\approx& \lambda^0_{\Gamma,\Gamma'}
+\delta\lambda_{i;\Gamma,\Gamma'}(t)\;.
\end{eqnarray}
For example, the first term on the r.h.s~of Eq.~(\ref{eq2}) becomes
 \begin{eqnarray}\nonumber
 \frac{\partial E^{\rm GA}(\tilde{\lambda}^{(*)},\tilde{\rho})}
{ \partial  \lambda_{i;\Gamma_1,\Gamma_2}^*}
 &\to&\sum_{k,\Gamma_3,\Gamma_4}
\frac{\partial^2 E^{\rm GA}(\tilde{\lambda}^{(*)},\tilde{\rho})}
{ \partial  \lambda^*_{i;\Gamma_1,\Gamma_2}
 \partial  \lambda_{k;\Gamma_3,\Gamma_4}}
\delta\lambda_{k;\Gamma_3,\Gamma_4}(t)\\
&+&\sum_{\upsilon,\upsilon'}
\frac{\partial^2 E^{\rm GA}(\tilde{\lambda}^{(*)},\tilde{\rho})}
{ \partial  \lambda^*_{i;\Gamma_1,\Gamma_2}
 \partial \rho_{\upsilon,\upsilon'}}
\delta \rho_{\upsilon,\upsilon'}(t)\;.
\end{eqnarray}
After the linearisation of Eqs.~(\ref{eq1}) and~(\ref{eq2}) and with the
 harmonic Ansatz
\begin{eqnarray}
\delta \lambda_{i;\Gamma,\Gamma'}(t)&=&
\delta \lambda_{i;\Gamma,\Gamma'}(\omega)e^{\rmi \omega t}\;,\\
\delta \rho_{\upsilon,\upsilon'}(t)&=&
\delta \rho_{\upsilon,\upsilon'}(\omega)
e^{\rmi \omega t}\;,
\end{eqnarray}
 we finally end up with a linear set of equations for 
 $\lambda_{i;\Gamma,\Gamma'}(\omega)$ and
$\delta\rho_{\upsilon,\upsilon'}(\omega)$
 which can be solved numerically.~
 Note that at zero frequency, the l.h.s.~of   Eq.~(\ref{eq2}) vanishes.~
 This equation then recovers exactly the `antiadiabaticity assumption'
 which was used in the previous TDGA formulation 
\cite{goe01,goe03,goe042,goe08}.~
Within the single band Hubbard model we will demonstrate in the 
following that the inclusion
of the full time-dependence of the variational
parameters $\lambda_{i;\Gamma,\Gamma'}(t)$ generates additional
features in the dynamical charge correlations which are 
absent in the `antiadiabatic approximation'.~

 \section{Response functions in the single-band Hubbard model}\label{hjud}
In the single-band Hubbard model the three most relevant response channels are
related to the coupling to time-dependent magnetic, charge and
pair fields.~This requires the computation of 
the (transversal) `magnetic
susceptibility'
\begin{equation}
\chi^{\rm T}_{i,j}(t)=
\langle  \langle
\hcd_{i,\uparrow}\hc_{i,\downarrow};\hcd_{j,\downarrow}\hc_{j,\uparrow}
  \rangle  \rangle  (t)\;,
 \end{equation}  
the `charge
susceptibility'
\begin{equation}
\chi^{\rm C}_{i,j}(t)=
\langle  \langle
\hat{n}_{i};\hat{n}_{j}
  \rangle  \rangle  (t)\;,
 \end{equation}
and the  `pairing
susceptibility' 
\begin{equation}
\chi^{\rm P}_{i,j}(t)=
\langle  \langle
\hcd_{i,\uparrow}\hcd_{i,\downarrow};\hc_{j,\downarrow}\hc_{j,\uparrow}
  \rangle  \rangle  (t)\;,
 \end{equation}  
or their respective Fourier transforms
\begin{equation}
\chi^{\rm T/C/P}(\vec{q},\omega)=\frac{1}{L}
\int_{-\infty}^{\infty}e^{\rmi \omega t}
 \sum_{i,j}
e^{\rmi (\vec{R}_i-\vec{R}_j) \vec{q}}
\chi^{\rm T/C/P}_{i,j}(t)\;,
 \end{equation}
where we have introduced $\hat{n}_{i}=\sum_{\sigma}\hat{n}_{i,\sigma}$.

\subsection{The magnetic and the pairing susceptibility}\label{dfg11}
If the state $|\Psi_{\rm S}\rangle$ is paramagnetic or is restricted to
longitudinal magnetic order, the charge and (transverse)
magnetic susceptibilities are decoupled.~
Moreover, if the ground state does not contain superconducting
correlations, also the pairing fluctuations are decoupled from
the magnetic and charge correlations.
In this case, the (mixed) second derivative of the energy 
 with respect to $\langle \hcd_{i,\uparrow}\hc_{i,\downarrow}   \rangle$ 
and $\lambda_{i;d}$ vanishes.
In a similar way, one can 
 show that the second derivative with respect to pairing fluctuations  
$\langle \hcd_{i,\uparrow}\hcd_{i,\downarrow}   \rangle$
 and  $\delta  \lambda_{i;d}$ vanishes.~Therefore, in both cases, 
the linearised 
differential equations~(\ref{eq1}) and~(\ref{eq2}) are decoupled
and the time dependence of $\delta  \lambda_{i;d}(t)$ does not 
enter the computation of (transverse) magnetic and pairing
correlation functions.
The susceptibilities are then solely determined by 
 the solution of Eq.~(\ref{eq1}) for the single-particle density matrix
and the present approach agrees with the previous formulation
of the TDGA involving the `antiadiabaticity approximation'
\cite{goe01,goe03,goe042,goe08}.
Therefore we concentrate in the following on the investigation
of the dynamical charge-charge correlation function where the
present approach is able to capture high-energy excitations
on the scale of the Hubbard repulsion due to the explicit
time dependence of the variational parameters.

\subsection{Dynamical charge susceptibility}\label{sec:dyn}
As derived in Sec. \ref{sec:oneb}, the time dependence of the
system is governed by small deviations of the density matrix 
$\delta\tilde\rho$, the double-occupancy
parameters $\delta D_i$ and the phase $\delta \eta_i$ 
from their saddle-point values (indicated by a `$0$' superscript).
Note that we consider a GA ground state  with $\eta_i^0=0$ 
such that $\delta \eta_i=\eta_i$.
The corresponding equations of motion
\begin{eqnarray}            
-\dot{\eta}_i &=& \frac{\partial E^{\rm GA}}{\partial \delta D_i}\;, \label{eq:dyn1} \\
\dot{\delta D}_i &=& \frac{\partial E^{\rm GA}}{\partial \eta_i} \;,
\label{eq:dyn2} \\
\rmi \delta\dot{\tilde{\rho}}&=&[\tilde{h}^{\rm GA},\delta\tilde{\rho}]\;,\label{eq:dyn3}
\end{eqnarray}
have to be solved for small deviations from the GA saddle-point.
For this purpose we expand the Gutzwiller energy functional 
Eq.~(\ref{eq:egahub}) up to second order in the fluctuations
\begin{equation}
E^{\rm GA} = E^{\rm GA,0}+ E^{\rm GA,1} + E^{\rm GA,2}
\end{equation}
around the saddle-point value $E^{\rm GA,0}$.

The first order contribution yields
\begin{eqnarray}
E^{\rm GA,1} &=& Tr\lbrace \tilde{h}^{\rm GA}\delta\tilde{\rho}\rbrace
+\sum_i \left(\frac{\partial E^{\rm GA}}{\partial D_i} \delta D_i 
+\frac{\partial E^{\rm GA}}{\partial \eta_i} \eta_i\right)
\end{eqnarray}
and the derivatives have to be taken at the saddle-point.
Here, the first term describes particle-hole excitations within the
`bare' Gutzwiller Hamiltonian. The second term $\sim \delta D_i$ vanishes
due to the saddle-point condition, whereas the last term can be expressed 
in the small amplitude limit as
\begin{equation}
\frac{\partial E^{\rm GA}}{\partial \eta_i} = \rmi\sum_\sigma\frac{q^0_{d,i,\sigma}}
{q^0_{\emptyset,i,\sigma}+q^0_{d,i,\sigma}}\left\lbrack \delta\rho_{i\sigma},\tilde{h}^{\rm GA}
\right\rbrack =  \sum_\sigma \frac{q^0_{d,i,\sigma}}
{q^0_{\emptyset,i,\sigma}+q^0_{d,i,\sigma}}\delta\dot{\rho}_{i\sigma}\;,
\end{equation}
where we have made use of Eq.~(\ref{eq:dyn3}).
Thus in the small amplitude limit it is convenient to introduce 
a 'displaced double occupancy' 
\begin{equation}
\tilde{D}_i = D_i - \sum_\sigma \frac{q^0_{d,i,\sigma}}
{q^0_{\emptyset,i,\sigma}+q^0_{d,i,\sigma}}\delta {\rho}_{i\sigma}
\end{equation}
such that the dynamics of $\eta_i$ and $\delta\tilde{D}_i$
can be expressed via the second order contribution of $E^{\rm GA}$ as
\begin{eqnarray}            
-\dot{\eta}_i &=& \frac{\partial E^{\rm GA,2}}{\partial \delta\tilde{D}_i}\;, \label{eq:dyn11} \\
\delta\dot{\tilde{D}}_i &=& \frac{\partial E^{\rm GA,2}}{\partial \eta_i} 
\label{eq:dyn22} \,.
\end{eqnarray}

The following evaluation of $E^{\rm GA,2}$ is exemplified for a homogeneous 
paramagnet but can be straightforwardly
generalised to arbitrary ground states.
In momentum space the second order expansion
involves fluctuations of  $\eta_q$ and $\delta \tilde{D}_q$ 
which are coupled to fluctuations of the local
$\delta\rho_q =\sum_{k,\sigma} \delta \rho_{k+q,k}^\sigma$ and
transitive $\delta T^+_{q} = \sum_{k,\sigma}\left\lbrack \varepsilon^0_{k+q} + \varepsilon^0_{k}\right\rbrack \delta \rho_{k+q,k}^\sigma$ charge densities:

\begin{eqnarray}                                                           
\delta E^{\rm GA,2}&=&\frac{1}{2N} \sum V_q \delta \rho_q \delta \rho_{-q}    
+ V^{T\rho}  
\frac{1}{N}\sum_q \delta T^+_q \delta \rho_{-q} \nonumber \\
&+& \frac{1}{\sqrt{N}}\sum_{q} g^{\tilde{D}\rho}_{q} \delta\tilde{D}_{-q} \delta\rho_{q} 
+ \frac{1}{\sqrt{N}}\sum_{q} g^{\tilde{D}T}_{q} \delta \tilde{D}_{-q} \delta T^+_{q} \nonumber \\
&+& \frac{1}{2} \sum_q K_q \delta \tilde{D}_q \delta \tilde{D}_{-q} 
+ \frac{1}{2} \sum_q \frac{1}{M} \eta_q \eta_{-q}
\label{equant}                                                           
\end{eqnarray}   
with
\begin{eqnarray}
V_q &=& \frac{1}{2} e_0q^0 (z_{++}''+ 2z_{+-}''+z_{--}'')                  
+ \frac{1}{2}(z'+z_{+-}')^2 \frac{1}{N}\sum_{k,\sigma} \varepsilon^0_{k+q}
\langle n_{k,\sigma}\rangle \nonumber \\
&+& 2\frac{q^0_{d}}{q^0_{\emptyset}+q^0_{d}}
\left( L_q + \frac{K_q}{2}\frac{q^0_{d}}{q^0_{\emptyset}+q^0_{d}}   \right)\;,\\
V^{T\rho}&=&\frac{1}{2}q^0(z'+z'_{+-}) + q^0 z_D'\frac{q^0_{d}}{q^0_{\emptyset}+q^0_{d}}\;,\\
g^{\tilde{D}\rho}_{q}&=& L_q + K_q \frac{q^0_{d}}{q^0_{\emptyset}+q^0_{d}} \;,\\
g^{\tilde{D}T}_{q}&=& q^0 z_D' \;, \\
K_q &=& 2 (z_D')^2 \frac{1}{N}\sum_{k,\sigma} \varepsilon^0_{k+q}
\langle n_{k,\sigma}\rangle \label{eq:kk} 
+ 2 \varepsilon^0q^0z_D'' \;,\\
\frac{1}{M} &=& -2 \varepsilon^0 q^0_{\emptyset} q^0_{d}\;,\label{eq:m}\\ 
L_q &=&  e_0q^0 (z_{+D}''+ z_{-D}'')
+ z_D' (z'+z_{+-}') \frac{1}{N}\sum_{k,\sigma} \varepsilon^0_{k+q}
\langle n_{k,\sigma}\rangle 
\end{eqnarray}
where at the saddle point the renormalisation factors become site and
spin independent (i.e. $q^0_{\emptyset,i,\sigma}=q^0_{\emptyset}$,
$q^0_{d,i,\sigma}=q^0_{d}$, $q^0_{i\sigma}=q^0$) and 
the primed letters denote derivatives which are specified in
\ref{ap:hop}.~We have also defined the 
non-renormalised single-particle
dispersion $\varepsilon^0_{k}$, whereas the GA quasi-particle dispersion
will be denoted as $\varepsilon_{k} = (q_{i,\sigma})^2 \varepsilon^0_{k}$.
Then from Eqs.~(\ref{eq:dyn11}),(\ref{eq:dyn22}) the
phase and double-occupancy dynamics is given by
\begin{eqnarray}
-\dot{\eta}_q &=& \frac{\partial E^{\rm GA,2}}{\partial \delta\tilde{D}_{-q}} \nonumber\\
&=& \frac{1}{\sqrt{N}} g^{\tilde{D}\rho}_{q}\delta\rho_{q}+
\frac{1}{\sqrt{N}} g^{\tilde{D}T}_{q}\delta T^+_{q}+K_q \delta{\tilde{D}}_q  
\label{eq:dyn111}\;, \\
\delta\dot{\tilde{D}}_q &=& \frac{\partial E^{\rm GA,2}}{\partial \eta_{-q}}=\frac{1}{M} \eta_q\;,
\label{eq:dyn222}
\end{eqnarray}
which upon Fourier transforming in the time domain yields
for the double occupancy
\begin{equation}
\tilde{D}_q= \frac{1}{\sqrt{N}} \left( 
\gamma^{\tilde{D}\rho}_{q}\delta\rho_{q} + \gamma^{\tilde{D}T}_{q}
\delta T^+_{q}\right){\cal D}^0(q,\omega) \;.
\end{equation}
Here we have defined the renormalised couplings
\begin{eqnarray}
\gamma^{\tilde{D}\rho}_{q} &=& \frac{g^{\tilde{D}\rho}_{q}}{\sqrt{2\Omega_q M}}\;, 
\label{eq:coupr}\\
\gamma^{\tilde{D}T}_{q} &=& \frac{g^{\tilde{D}T}_{q}}{\sqrt{2\Omega_q M}}\label{eq:coupt}\;,
\end{eqnarray}
and the `double occupancy' Green's function
\begin{equation}\label{eq:ddgf}
{\cal D}^0(q,\omega) = \frac{2 \Omega_q}{\omega^2 -\Omega_q^2}\;,
\end{equation}
which has poles at $\Omega_q^2=K_q/M$.
In case of a half-filled 
system one obtains
\begin{equation}\label{eq:omq}
\Omega_q^2=16 \varepsilon_0^2 \left\lbrack 1 - u^2 \kappa_q\right\rbrack\;, 
\end{equation}
where $\varepsilon_0$ is the energy of the non-interacting system,
$u\equiv U/|8\varepsilon_0|$, and
$\kappa_q = \frac{1}{\cal D} \sum_{i=1}^{\cal D} \cos(q_i)$ .
Interestingly the Brinkman-Rice transition $u=1$ can therefore
be associated with a `soft mode' where $\Omega_{q=0} \to 0$. This
softening is clearly associated to the extra gauge invariance
condition that appears at the  Brinkman-Rice point which makes the
mass $M$ to diverge. 

Equations (\ref{equant})-(\ref{eq:dyn222}) show that in the 
 present approach the interacting electron problem can be mapped to an
effective electron-boson problem. Electron quasiparticles are coupled
to a boson representing fluctuations of the double occupancy
(``doublon fluctuations'') or of its conjugate variable. 
Notice that the mass $M$ of the fluctuation field diverges in the two
situations discussed after Eq.~(\ref{eq:lagran1b}) either
$E\rightarrow 0$ of $D\rightarrow 0$ (c.f
Eqs.~(\ref{eq:qempty}),(\ref{eq:d}) and (\ref{eq:m}). This follows from the fact that 
in those cases $\eta$ becomes a gauge phase and therefore gauge 
invariance requires that the last term of Eq.~(\ref{equant}) vanishes.

In analogy with electron-phonon problems the dressed fluctuations are
most conveniently evaluated 
by defining a (non-interacting) susceptibility matrix
\begin{eqnarray}
\underline{\underline{\chi^{ee,0}_{q}(\omega)}}&=&
\frac{-1}{N}\int_0^\beta d\tau \mbox{e}^{i\omega_n \tau}
\left(\begin{array}{cc}
\langle {\cal T} \delta\rho_\qvec(\tau)\delta\rho_{-\qvec}(0)\rangle
& \langle {\cal T} \delta \rho_\qvec(\tau)\delta T^+_{-\qvec}(0)\rangle \\
\langle {\cal T} \delta T^+_\qvec(\tau)\delta\rho_{-\qvec}(0)\rangle
& \langle {\cal T} \delta T^+_\qvec(\tau)\delta T^+_{-\qvec}(0)\rangle \\
\end{array}\right) \nonumber \\
&=&\frac{1}{N} \sum_{k\sigma}
\left(\begin{array}{ccc}
1 & \epsilon^0_{\kvec+\qvec}+ \epsilon^0_\kvec \\
\epsilon^0_{\kvec+\qvec}+\epsilon^0_\kvec & 
\left(\epsilon^0_{\kvec+\qvec}+ \epsilon^0_\kvec\right)^2 \\ 
\end{array}\right)
\frac{n_{k+q,\sigma} - n_{k\sigma}}
{\omega+\epsilon_{k+q}-\epsilon_{k}} \label{eq:chi}
\end{eqnarray}
and an effective interaction matrix which is composed of the bare
electron-electron interaction and the second order `bosonic
contribution' 
\begin{displaymath}
\underline{\underline{\widetilde{V}^{ee}_{q}(\omega_n)}} =
\left(\begin{array}{cc}
V_q & V^{T\rho} \\
V^{T\rho} & 0 
\end{array}\right) +
\left(\begin{array}{cc}
(\gamma^{\tilde{D}\rho}_q)^2 & \gamma^{\tilde{D}\rho}_q g^{AT}_q\\
\gamma^{\tilde{D}\rho}_q \gamma^{\tilde{D}T}_q & (\gamma^{AT}_q)^2
\end{array}\right) {\cal D}^0(q,\omega) \, .
\end{displaymath}
The susceptibility for the interacting system is then obtained 
from the following RPA series    
\begin{equation}
\label{eqrpa}   
\underline{\underline{\chi_{q}}} = \underline{\underline{\chi^0_q}} 
+\underline{\underline{\chi^0_q}}\,
\underline{\underline{\widetilde{V}^{ee}_{q}}}       
\underline{\underline{\chi_{q}}}
\end{equation}
Note that in the static limit $\omega \to 0$ 
the matrix $\underline{\underline{\widetilde{V}^{ee}_{q}(0)}}$
is exactly the effective
interaction obtained within the antiadiabaticity condition in 
Refs.~\cite{goe01,goe03}.

\section{Results}\label{sec:res}
In this section, we present results for the local dynamical
charge correlation function 
\begin{equation}\label{eq:chcorr}
\chi_{\rm loc}(\omega)=\sum_{q\ne 0} \frac{|\langle 0|\hat{n}_i|q\rangle|^2}{\omega+E_q-E_0-\rmi\eta}
\end{equation}
where $|0\rangle$ and $|q \rangle$ denote ground and excited states
of the single-band Hubbard model.
We first study the low-density regime where
we compare the TDGA spectra with exact results for the case of two particles.
For higher densities we study the performance of the TDGA 
by comparing with DMFT and exact diagonalisation results.

\subsection{Low-density regime}
In the low-density limit $n\to 0$ the energy of the non-interacting system
is determined by $\varepsilon_0=-B n$ where $2B$ would be the total
bandwidth in case of a rectangular density of states.
The saddle point double occupancy in this limit reads as 
\begin{equation}
D_{n\to 0}^0=\frac{n^2}{4}\frac{1}{\left(1+\frac{U}{2B}\right)^2}
\end{equation}
which allows for the computation of the frequency of double-occupancy 
oscillations as $\Omega_q = 2 B +U$, i.e., it is independent of momentum.
Since the coupling to these fluctuations Eqs.~(\ref{eq:coupr}),(\ref{eq:coupt}) 
vanishes
with $n$  we expect the local TDGA charge correlations to be composed
of a renormalised low-energy part for $0 < \omega < 2B$ and a 
high-energy excitation at $\omega = 2 B +U$ with spectral weight
proportional to the density.

In case of two particles one can determine the eigenenergies  in
Eq.~(\ref{eq:chcorr}) from \cite{long}
\begin{equation}\label{eq:exc}                 
\frac{1}{U}=\frac{1}{N}\sum_k\frac{1}{E-\epsilon_k-\epsilon_{k+q}}  
\end{equation}                                                      
where $-4t \le \epsilon_k \le 4t$ denotes the single-particle dispersion
with bandwidth $2B=8t$.~The ground state is obtained for $q=0$ and
both particles at the bottom of the band, i.e.,~$E_0 \approx -2B$.
A particular solution in Eq.~(\ref{eq:exc}) is obtained for
${\bf Q}={\bf q}=(\pi,\pi)$ at $E_{\bf Q}=U$ so that the excited state
$E_{\bf Q}-E_0=2 B +U$ corresponds to our TDGA result discussed above.~
In addition, the exact solution involves two-particle excitations which are
not present in the TDGA.~The maximum excitation energy is obtained
for $q=0$ which can be estimated
for a rectangular density of states as $\omega= 4B/(e^{1/U}-1)$.
The weight of these excitations in Eq.~(\ref{eq:exc}) vanishes   
for zero momentum transfer but clearly the exact solution displays high
energy features in addition to the TDGA for small ${\bf q}$ 
due to the coupling between particle-hole and particle-particle excitations.

\begin{figure}[h]
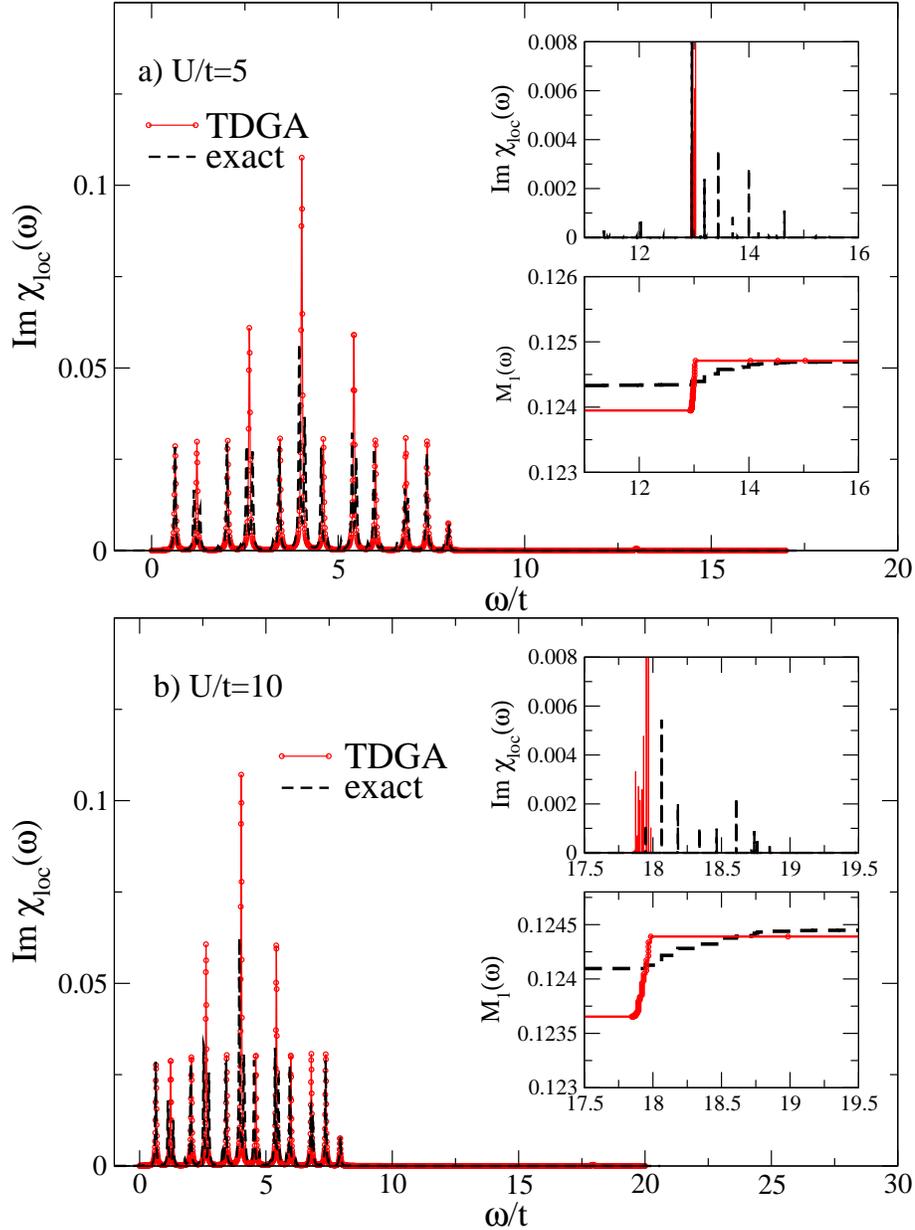

\centering
\includegraphics[width=12cm,clip=true]{n88u5.eps}\\
\includegraphics[width=12cm,clip=true]{n88u10.eps}
\caption{Local charge susceptibility for $2$ particles
on a $8\times 8$ lattice with (a) $U/t=5$ and (b) $U/t=10$.
We compare spectra for the exact result (black, dashed) with those
obtained within the TDGA (solid, circles).
The upper insets show the high-energy part of the spectra.      
Broadening of the excitations is $\epsilon=0.02t$ (main panels)
and $\epsilon=10^{-4}t$ (upper insets).~The lower insets depict
the high-energy part of the first moments $M_1(\omega)$.}      
\label{fig1}                                                   
\end{figure}

Figure~\ref{fig1} displays  the low-density local charge susceptibility
Eq.~(\ref{eq:exc}), evaluated with TDGA and exact diagonalisation
for $U/t=5$ and $U/t=10$, respectively.
Results have been obtained for       
$2$ particles on a $8\times 8$ square lattice lattice (only nearest 
neighbor hopping $-t$).~

As anticipated above, the spectra consist of the (dominant) 
low-energy bandlike particle-hole excitations in the range 
$0<\omega<2B$ (cf.~main panels) and a high-energy part at
$\omega \approx 2B+U$ which is resolved in the upper insets.
Similar to the previous investigations, which were
based on the `antiadiabaticity assumption' \cite{goe01,goe03},
the TDGA gives a very good account of the low-energy part with
respect to both, energy and spectral weight of the excitations.
The new feature, which was previously missing \cite{goe01,goe03}
in the TDGA, is the high-energy part due to the explicit consideration
of the double-occupancy time dependence. 
In order to estimate the associated spectral weight, we show in the
lower insets to Fig.~\ref{fig1} the first moment
\begin{equation}\label{eq:fmom} 
M_1(\omega)=\int_0^\omega\!\! d\nu\, \nu\, \chi_{\rm loc}(\nu)\;,
\end{equation}
which fulfills the sum rule 
\begin{equation}
M_1(\infty)=-\langle T \rangle \;.
\end{equation}
Here $\langle T \rangle$ denotes the kinetic energy which in
case of the TDGA has to be evaluated on the basis of the GA.
From Fig.~\ref{fig1} it turns out that the onset of the 
high-energy excitations is accurately captured by the TDGA, however,
it overestimates the associated spectral weight. 
On the other hand, this `additional' weight is partially compensated
for in 
the bandlike excitations such that the kinetic energy of the Gutzwiller 
approximation (i.e.~$M_1(\infty)$) again gives a good account
of the exact result.~

\subsection{Density dependence}

\begin{figure}[htb]
\includegraphics[width=0.8\textwidth,angle=-90,clip=true]{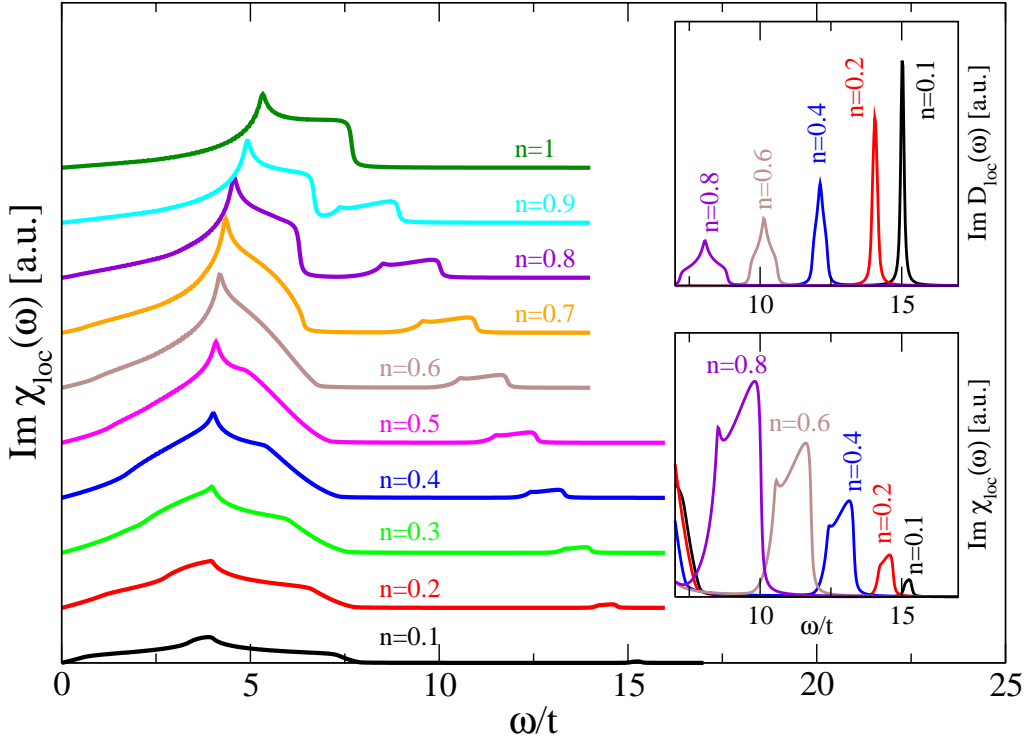}
\caption{TDGA local charge susceptibility for a two-dimensional
square lattice Hubbard model ($U/t=8$, nearest-neighbor hopping $-t$) 
at different densities. The inset displays the high-energy
part on an enlarged scale whereas the upper inset shows the imaginary
part of the double-occupancy propagator.}                                     
\label{fig2}                                                               
\end{figure}

We proceed by studying the doping dependence of $\chi_{\rm loc}(\omega)$
as a function of doping which is shown in Fig.~\ref{fig2} for
$U/t=8$ for a square lattice.
The spectra again separate into
low-energy band-like excitations and a high-energy part due to
the double-occupancy time dependence.
Starting form the low density limit the 
overall weight of the high-energy excitations
increases approaching half filling. In addition 
the high-energy feature 
shifts to lower frequencies upon doping as shown in the lower inset to 
Fig.~(\ref{fig2}).

We show also in the upper inset the 
imaginary part of the local double-occupancy propagator, i.e.
\begin{equation}
{\mbox{Im}} D_{\rm loc}^0(\omega)={\mbox{Im}}\frac{1}{N}\sum_q {\cal D}^0(q,\omega)
\end{equation}
and ${\cal D}(q,\omega)$ has been defined in Eq.~(\ref{eq:ddgf}).

The double-occupancy excitations evolve from 
$\Omega=2B+U$ ($=16t$ for the present parameters) in the limit $n\to 0$ 
to the frequencies $\Omega_q$ defined in Eq.~(\ref{eq:omq}) 
for the case of half-filling.
In order to understand the doping dependence of the high-energy
feature one has also to take into account the couplings 
Eqs.~(\ref{eq:coupr}),(\ref{eq:coupt}).
The coupling to the local fluctuations, $\gamma_q^{\tilde{D}\rho}$ 
[Eq.~(\ref{eq:coupr})] is continuously
decreasing with the charge carrier concentration and
dominates at all dopings except close to half-filling
where it vanishes. On the other hand, the coupling
to transitive fluctuations $\gamma_q^{\tilde{D}T}$ [Eq.~(\ref{eq:coupt})]
is significantly smaller and only weakly doping dependent.
Since at half-filling the coupling between local and transitive 
fluctuations vanishes ($V^{T\rho}=0$), the local charge correlations are only
renormalised by the static density-density interactions $V_q$.
Thus at exactly half-filling the coupling to the double-occupancy
fluctuations vanishes and the $n=1$ curve in Fig.~(\ref{fig2})
corresponds to the `antiadiabaticity' result derived in
Refs.~\cite{goe01,goe03}. 
With decreasing doping the increasing coupling between local
density and double fluctuations increases and induces the
shift of the high-energy feature to large frequencies. 

In order to check the quality of the TDGA at larger doping {\it vs.}
other approaches we compare our results with dynamical mean-field
theory (DMFT)\cite{dmft}.
Despite freezing the spatial fluctuations beyond mean-field, DMFT
takes fully into account the local quantum dynamics and it is in
particular reliable to describe the evolution of the spectral weight as
a  function of temperature and other control parameters like doping.
DMFT maps the lattice model onto an Anderson Impurity model, which is
solved with an `impurity solver', which in the present work is
exact diagonalisation \cite{ed-dmft}. Within this method the bath of
the AIM is discretised into $N_b$ levels, which here is taken to be
9. 
A Matsubara grid defined by an effective temperature $\beta =80/t$ is
used and the stability of the results as a function of the two control
parameters has been checked.  
Within DMFT, the local dynamical correlation functions can be obtained
without further approximation. As a result of the discrete bath, the 
spectra appear more spiky than in the actual solution, but it has 
been shown that this approach describes accurately the evolution of
the spectral weight (for instance of the optical conductivity) as a function of the various control 
parameters \cite{Toschi_PRL,Nicoletti_PRL}.

\begin{figure}[htb]
\includegraphics[width=0.8\textwidth,clip=true]{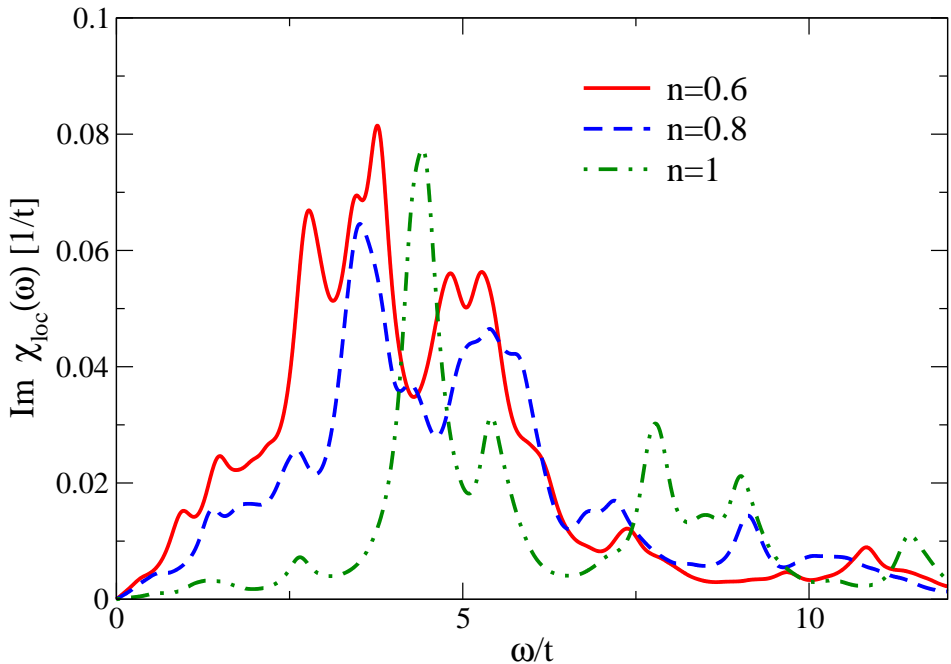}
\caption{DMFT local charge susceptibility for a two-dimensional
square lattice Hubbard model ($U/t=8$, nearest-neighbor hopping $-t$) 
at densities $n=0.6$, $n=0.8$, and $n=1$.}                                     
\label{fig3}                                                               
\end{figure}                                                               

Figure \ref{fig3} (main panel) shows the local charge susceptibility obtained
within DMFT for different fillings and $U/t=8$. Despite the peaky
structure (which hampers a direct comparison with the TDGA results) 
it is obvious that the main features correspond to those
obtained within the TDGA, i.e., bandlike excitations on the scale
of $8t$ and additional higher energy excitations which soften and
gain spectral weight upon increasing density and approaching the
insulating phase.

\begin{figure}[htb]
\includegraphics[width=0.8\textwidth,clip=true]{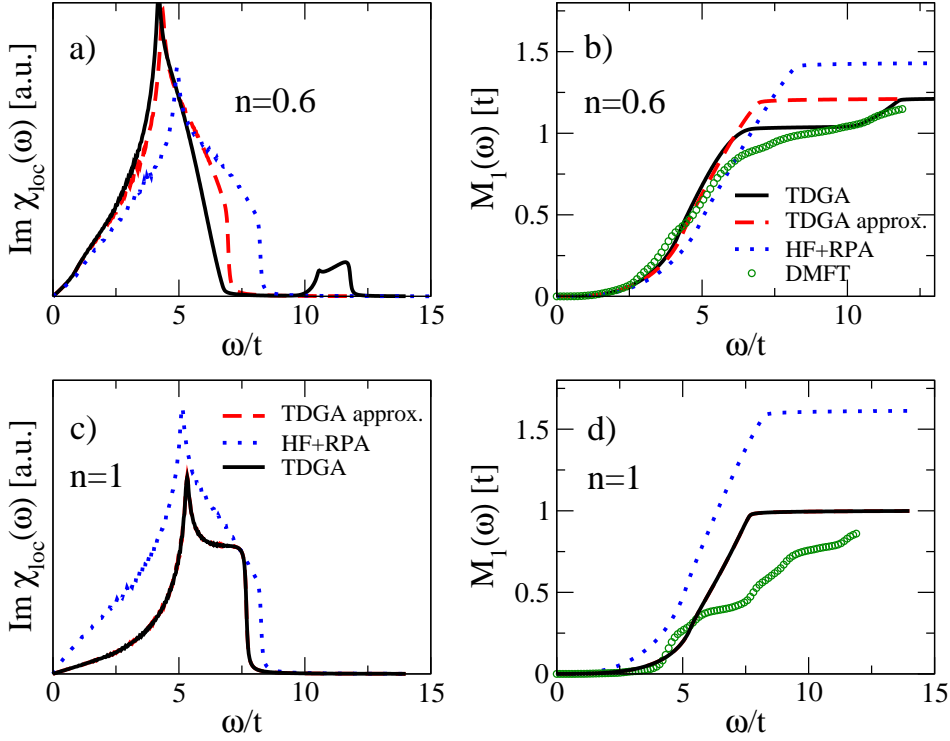}
\caption{Imaginary part of the local charge susceptibility at
densities $n=0.6$ (panel a) and $n=1$ (panel c) evaluated
with the TDGA (solid black), TDGA with antiadiabaticity (dashed red),
and HF+RPA (dotted blue). Panels c,d report the corresponding
evolution of the first moment Eq.~(\ref{eq:fmom}) where also
the results from DMFT (circles, green) are shown.
Onsite repulsion: $U/t=8$.}                                     
\label{fig2.5}                                                               
\end{figure}

In order to test the performance of the TDGA 
we show in Fig.~\ref{fig2.5} (panels b,d) 
a comparison of the first moment $M_1(\omega)$,
Eq.~(\ref{eq:fmom}), evaluated within TDGA (black solid lines) 
and DMFT (green circles) for $U/t=8$ and densities $n=0.6$ (panel b)
and $n=1.0$ (panel d). 
As anticipated above the TDGA gives a rather good
account of the spectral weight evolution at lower densities
where it is in good agreement with the DMFT data (Fig.~\ref{fig2.5}, panel b)
given the uncertainties 
due to the finite number of bath states.
However, due to the vanishing coupling between electrons and
double-occupancy fluctuations at half-filling, all the TDGA
spectral weight is contained in the band-like excitations
in this limit so that the `antiadiabaticity' result
of Refs.~\cite{goe01,goe03} is recovered. 
Therefore the corresponding first moment 
 increases much faster than DMFT but nevertheless both moments approach
for $\omega\to\infty$ due to the agreement in the kinetic energies.
One should note that at half-filling the GA ground state
actually corresponds to a spin-density wave and also for such
symmetry-broken states we find that at half-filling the
antiadiabaticity result of Refs.~\cite{goe01,goe03} is valid.
In fact, the TDGA charge excitations on top of such a ground state
are in reasonable agreement with exact diagonalisation
as demonstrated in Fig.~5 of Ref.~\cite{goe03}.

For comparison, we also show in Fig.~\ref{fig2.5} the 
result of HF+RPA theory together with the spectra of our previous 
`approximate' TDGA \cite{goe01,goe03} where the double-occupancy fluctuations
have been antiadiabatically eliminated.
As discussed above, at half-filling the antiadiabatic approximation 
agrees with the `exact' evaluation of the TDGA
(cf. panels c,d in Fig.~\ref{fig2.5}). 
The difference becomes pronounced at lower doping where the
high-energy feature gets significant spectral weight and due to
repulsion induces a softening of the bandlike excitations.
From panel a of Fig.~\ref{fig2.5} it is clear that both effects
are essential for reproducing the very good agreement of the
first moment with the DMFT result at $n=0.6$ whereas the approximate TDGA 
interpolates the spectral weight between the bandlike
and high-energy excitations.
Note, however, that for small frequencies $\omega \to 0$ the approximate
result agrees with the `exact' TDGA as has already been discussed in 
Sec. \ref{sec:sma}. In addition, the first moment
of both approaches agrees in the limit $\omega \to \infty$ since
it is set by the static GA kinetic energy.

In case of the conventional HF+RPA approach, where the
local charge susceptibility is given by
\begin{equation}
\chi_{\rm loc}^{\rm HF+RPA}(\omega)=\frac{1}{N}\sum_q\frac{\chi^0_q(\omega)}
{1-\frac{U}{2}\chi^0_q(\omega)}
\end{equation} 
and $\chi^0_q(\omega)$ has the same structure than the $(11)$-element
of Eq.~(\ref{eq:chi}) but with non-renormalised single-particle
dispersions $\varepsilon_k^0$.
Since HF+RPA obviously does not capture the double-occupancy dynamics,
the local charge fluctuations originate from the band-type
excitations which are renormalised due to the RPA denominator
and high frequency excitations are absent.
Moreover, HF overestimates the kinetic energy so that the first moment
of $\chi_{\rm loc}^{\rm HF+RPA}$ overshoots both the TDGA and DMFT result.
As can be seen from  Fig.~\ref{fig2.5} this discrepancy
is most apparent close to half-filling where the renormalisation of
the kinetic energy due to correlation effects is more pronounced.
Upon reducing the band filling the low-energy part of the 
HF+RPA spectra approaches the TDGA result where, however, the latter
approach additionally captures the high frequency excitations at $2B+U$
with spectral weight $\sim n$.

\section{Conclusions}\label{sec:con}
In this paper we have developed the time-dependent Gutzwiller approximation
for multi-band Hubbard models. This approach is based on a time-dependent
variational principle where expectation values are evaluated with
the Gutzwiller variational wave-function in the limit of
infinite dimensions.~In contrast to the standard Gutzwiller approximation
\cite{GUTZ1,GUTZ2,GEBHARD} both, variational parameters and the underlying
Slater determinant, acquire a time dependence. In this regard our
calculations generalise earlier investigations by  
Schir\'{o} and Fabrizio \cite{schiro1,schiro2} who have studied quantum
quenches in homogeneous systems where the time dependence of the
density matrix does not couple to that of the variational parameters.
On the other hand, momentum (or space) dependent out-of equilibrium
displacements of the system require such a coupling as evident
from our generalised equations of motion Eqs.~(\ref{eq1}),(\ref{eq2}).

We have applied this theory in the small-amplitude, i.e., 
linear-response limit and exemplified for the case
of dynamical charge correlations in the single-band Hubbard model.
In an earlier formulation of the TDGA the so-called 
`antiadiabaticity approximation' \cite{goe01,goe03}
has been applied, where the dynamics of the double-occupancy
parameters was slaved by that of the density matrix.~
In contrast,
the present approach explicitly incorporates the time dependence
of the double-occupancy variational parameters which 
agrees with the previous formulation in the static limit.
In addition it improves the theory in Refs.~\cite{goe01,goe03}
by incorporating the high-energy features which are on the
scale of the Hubbard repulsion for small densities and 
which position is in good agreement with that of
exact diagonalisation. On the other hand, the spectral weight of the
high-energy excitations 
is overestimated within the TDGA although it significantly improves
the standard HF+RPA approach in this regard.~Further refinement of the
theory could be achieved by including the coupling between particle-hole
and particle-particle excitations which have been studied in 
Ref.~\cite{goe08} in the framework of the GA.~

It is interesting that in the present approach the Brinkman-Rice
 transition appears signaled by a collective mode whose frequency goes to
zero. This is not due to the doublon fluctuation stiffness 
 becoming soft but because the mass of the fluctuations
diverges. We have shown that this divergence appears each time the
double occupancy becomes a conserved quantity which is the case in the 
 Brinkman-Rice case where $D=0$. It remains to be seen which of these
 feature remain in an exact description although the similarity of the
 TDGA results with dynamical mean field theory (DMFT) suggest that at least
in an  approximate way this physics survives in real Mott transitions. 

Within the DMFT it is quite diffult to study systems in which the momentum 
dependence of collective excitations is important 
as, for example, spin waves in insulators \cite{lorenzana34}. 
In such cases, the 
 TDGA provides us with an important additional tool which complements
 the DMFT.

\ack
\section{Acknowledgments}
M.~C.~is financed by EU/FP7 through ERC Starting Grant SUPERBAD, Grant 
Agreement 240524, and GO FAST, Grant Agreement 280555. J. L. is 
 supported  by the Italian Institute of Technology
through the project NEWDFESCM.

\appendix         
\section{Physical meaning of the phases}\label{ap:phase}
To understand the physical meaning of the 
phases appearing in Eq.~(\ref{eq:renormeta}) and how they affect the
hopping amplitude consider the following two-site
example in which the non interacting state is uniform,
$$|\Psi_{\rm S}\rangle=\frac12(\hat{c}_{1,\uparrow}^{\dagger}+\hat{c}_{2,\uparrow}^{\dagger})(\hat{c}_{1,\downarrow}^{\dagger}+\hat{c}_{2,\downarrow}^{\dagger})|{\rm
vac}\rangle $$
but the projectors are given by Eq.~(\ref{pi_one}) with all
$\lambda_\Gamma=1$ except $\lambda_{2d}=-1$ which corresponds to
$\eta_2=\pi$ and the other phases zero leading to
$q_{2,\sigma}=0$, i.e., destructive interference.  
The projected wave function reads, 
$$|\Psi_G\rangle=\frac12(\hat{c}_{1,\uparrow}^{\dagger}\hat{c}_{1,\downarrow}^{\dagger}
- \hat{c}_{2,\uparrow}^{\dagger}\hat{c}_{2,\downarrow}^{\dagger}+\hat{c}_{1,\uparrow}^{\dagger}\hat{c}_{2,\downarrow}^{\dagger}+\hat{c}_{2,\uparrow}^{\dagger}\hat{c}_{1,\downarrow}^{\dagger})|{\rm
vac}\rangle\;. $$ 
The exact off-diagonal density matrix in the Gutzwiller
wave function is given by the overlap
between the states,
\begin{eqnarray*}
  \hat{c}_{1,\uparrow}|\Psi_G\rangle&=&\frac12(\hat{c}_{1,\uparrow}^{\dagger}+
\hat{c}_{2,\uparrow}^{\dagger})|{\rm
    vac}\rangle\;,\\
 \hat{c}_{2,\uparrow}|\Psi_G\rangle&=&\frac12(\hat{c}_{1,\uparrow}^{\dagger}-
\hat{c}_{2,\uparrow}^{\dagger})|{\rm 
   vac}\rangle\;.
\end{eqnarray*}
We see that in this zero dimensional example the `background' electron
remains with the `wrong' phase (or momentum)
leading  to zero overlap, in accord with the GA derived in infinite dimensions.
Notice, however, that if also $\lambda_{1d}=-1$ the overlap is
finite in the exact evaluation while it is zero in the GA. 
Clearly such kind of process depends on the correlation
between the phases on different sites  and can not be captured by the
factorised form $\sim q_{1,\sigma}^*q_{2,\sigma}$ of the GA.

\section{Derivatives of the renormalisation factors}\label{ap:hop}
In Sec.~\ref{sec:dyn} we have introduced derivatives of the 
hopping renormalisation factor $q_{i,\sigma}$, evaluated at the saddle-point
of a paramagnetic system.~These are defined as follows:

\begin{eqnarray*}
&&\frac{\partial q_{i \sigma}}{\partial \rho_{ii 
\sigma}}\equiv z^{'},  \ \ \  \frac{\partial q_{i\sigma}}{\partial \rho
_{ii -\sigma}}\equiv z^{'}_{+-}, \                                              
\frac{\partial q_{i \sigma}}{\partial D_{i}}\equiv z^{'}_{D}                    
\label{z1} \\                                                                   
&& \frac{\partial^2 q_{i \sigma}}{\partial \rho^{2}_{ii \sigma}} \equiv z^{''}_{++},\
\frac{\partial^2 q_{i \sigma}}{\partial \rho_{ii \sigma}\partial \rho_{ii -\sigma}}
\equiv z^{''}_{+-}, \ \frac{\partial^2 q_{i \sigma}}{\partial \rho^{2}_{ii -\sigma}} 
\equiv z^{''}_{--}  \nonumber \\                                           
&&  \frac{\partial^2 q_{i \sigma}}{\partial D^{2}_{i}}\equiv z^{''}_{D}, \ 
\frac{\partial^2 q_{i \sigma}}{\partial \rho_{ii \sigma}\partial D_{i}}
\equiv z^{''}_{+D}, \ \frac{\partial^2 q_{i \sigma}}{\partial \rho_{ii -\sigma}
\partial D_{i}}\equiv z^{''}_{-D}       
\label{z2}                                                                      
\end{eqnarray*}


\section*{References}

\begin{thebibliography}{9}
\bibitem{conte} S.~D.~Conte, C.~Giannetti, G.~Coslovich, F.~Cilento,
D.~Bossini, T.~Abebaw, F.~Banfi, G.~Ferrini, H.~Eisaki,
M.~Greven, A.~Damascelli, D.~v.~d.~Marel, and G.~Parmigiani, Science {\bf 335},
1600 (2012).
\bibitem{fausti} D.~Fausti, R.~I.~Tobey, N.~Dean, S.~Kaiser, A.~Dienst, M.~C.
Hoffmann, S.~Pyon, T.~Takayama, H.~Takagi, and A.~Cavalleri,
Science 331, 189 (2011).
\bibitem{mansart} B.~Mansart, J.~Lorenzana, A.~Mann, A.~Odeh, 
M.~Scarongella, M.~Chergui, and F.~Carbone, arXiv:1112.0737.
\bibitem{eckstein} M.~Eckstein and P.~Werner, Phys.\ Rev.~B {\bf 84}, 035122 (2011).
\bibitem{goe01} G.~Seibold and J.~Lorenzana, Phys.\ Rev.~Lett.~{\bf 86}, 2605 (20
01).                                                                           
\bibitem{goe03} G.~Seibold, F.~Becca, and J.~Lorenzana, Phys.\ Rev.~B {\bf 67},
085108 (2003).
\bibitem{goe042} G.~Seibold, F.~Becca, P.~Rubin, and                            
                 J.~Lorenzana, Phys.\ Rev.~B {\bf 69}, 155113 (2004).          
\bibitem{goe08} G.~Seibold, F.~Becca, and J.~Lorenzana,                         
                Phys.\ Rev.\ Lett.~{\bf 100}, 016405 (2008);                      
                G.~Seibold, F.~Becca, and J.~Lorenzana,                         
                Phys.\ Rev.~B {\bf 78}, 045114 (2008).                          
\bibitem{GUTZ1} M.C.~Gutzwiller, Phys.\ Rev.\ Lett.~{\bf 10}, 159 (1963).        
\bibitem{GUTZ2} M.C.~Gutzwiller, Phys.\ Rev.~{\bf 134}, A 923 (1964);            
    {\it ibid.} {\bf 137}, A 1726 (1965).                                      
\bibitem{GEBHARD} F.~Gebhard, Phys.\ Rev.~B {\bf 41}, 9452 (1990).
\bibitem{oelsen1} E.~v.~Oelsen, G.~Seibold, and J.~B\"{u}nemann, 
                  New Journal of Physics {\bf 13}, 113031 (2011).
\bibitem{oelsen2} E.~v.~Oelsen, G.~Seibold, and J.~B\"unemann, 
                  Phys.\ Rev.\ Lett.~{\bf 107}, 076402 (2011).
\bibitem{ring} P.~Ring and P.~Schuck,                                           
                {\em The nuclear many-body problem} (Springer-Verlag,           
                New York, 1980).                                              
\bibitem{blaizot} J.~Blaizot, G.~Ripka {\it Quantum theory of finite systems},  
                  MIT Press, 1986.                                            
\bibitem{goe12} G.~Seibold, M.~Grilli, and J.~Lorenzana, Physica C {\bf 481},
                132 (2012).
\bibitem{lor03} J.~Lorenzana and G.~Seibold, Phys.\ Rev.\ Lett.~{\bf 89}, 136401 (
2002).                                                                         
\bibitem{goetz05} G.~Seibold and J.~Lorenzana, Phys.\ Rev.\ Lett.~{\bf 94},       
107006 (2005).                                                                 
\bibitem{sei06} G.~Seibold and J.~Lorenzana, Phys.\ Rev.~B {\bf 73},              
144515 (2006).                                                                 
\bibitem{ugenti10} S.~Ugenti, M.~Cini, G.~Seibold, J.~Lorenzana, E.~Perfetto, an
d G.~Stefanucci, Phys.\ Rev.~B {\bf 82}, 075137 (2010).                         
\bibitem{schiro1} M.~Schir\'{o} and M.~Fabrizio, Phys.\ Rev.\ Lett.~{\bf 105},    
                 076401 (2010).                                               
\bibitem{schiro2} M.~Schir\'{o} and M.~Fabrizio, Phys.\ Rev.~B {\bf 83},         
                 165105 (2011).                                               
\bibitem{fabrizio} M.~Fabrizio, arXiv:1204.2175.                             
\bibitem{andre} P.~Andr\'e, M. Schir\`o, and M.~Fabrizio, Phys.\
  Rev.~B {\textbf{85}}, 205118 (2012)
\bibitem{mazza} G.~Mazza and M.~Fabrizio, arXiv:1210.2034
\bibitem{kramers} P. Kramers and M. Saraceno (eds.), 
{\it Geometry of the Time-Dependent Variational Principle in Quantum Mechanics},
Lecture Notes in Physics {\bf 140}, Springer Berlin Heidelberg (1981). 
\bibitem{buenemann1998} J.~B\"unemann, W.~Weber, and F.~Gebhard, 
                   Phys.\ Rev.~B {\bf 57}, 6896 (1998).                     
\bibitem{buenemann2005}
 J.~B{\"u}nemann, F.~Gebhard,  and  W.~Weber,
 in: Frontiers in Magnetic Materials, edited by A.~Narlikar,  (Springer,
  Berlin, 2005).
 \bibitem{buenemann2011d} J.~B\"unemann, T.~Schickling, and F.~Gebhard,
 Europhys.\ Lett.~{\bf 98}, 27006 (2012). 
\bibitem{buenemann2013} J.~Kaczmarczyk, J.~Spa\l ek, T.~Schickling, and 
 J.~B\"unemann (unpublished).
   \bibitem{buenemann2012} J.~B\"unemann, T.~Schickling, F.~Gebhard, and 
 W.~Weber, Phys.\ Status Solidi~{\bf 249}, 1282 (2012).     
\bibitem{long} M.~W.~Long and R.~Fehrenbacher, J.~Phys.: Condens.\
  Matter.~{\bf 2}, 10343 (1990).    
\bibitem{dmft} A.~Georges, G.~Kotliar, W.~Krauth, and M.~J.~Rozenberg,
 Rev. Mod. Phys. \textbf{68}, 13 (1996)
\bibitem{ed-dmft} M. Caffarel and W. Krauth, Phys. Rev. Lett. \textbf{72}, 1545
(1994); M. Capone, L. de' Medici, and A. Georges, Phys. Rev. B
\textbf{76}, 245116 (2007).
Rev.~Mod.~Phys.~{\bf 68}, 13 (1996).
\bibitem{Toschi_PRL} A.~Toschi, M.~Capone, M.~Ortolani, P.~Calvani,
  S.~Lupi, and C.~Castellani, Phys. Rev. Lett. \textbf{95}, 097002
  (2005).
\bibitem{Nicoletti_PRL} D.~Nicoletti, O.~Limaj, P.~Calvani,
  G.~Rohringer, A.~Toschi, G.~Sangiovanni, M.~Capone, K.~Held, S.~Ono,
  Y.~Ando, and S.~Lupi, Phys. Rev. Lett. \textbf{105}, 077002 (2010).
\bibitem{lorenzana34} J.~Lorenzana, G. Seibold, and R. Coldea, 
Phys.\ Rev.~B {\bf 72},              
224511 (2005).       
\end{thebibliography}
\end{document}